\documentclass[pra,amsmath,amssymb,showpacs,superscriptaddress,twocolumn]{revtex4}
\input epsf.tex
\usepackage{graphicx}
\usepackage{amsthm}
\usepackage{array}
\usepackage{longtable}
\usepackage{dcolumn}
\usepackage{bm,bbm}
\usepackage[usenames,dvipsnames]{color}
\usepackage{amsmath}
\usepackage[hypertex]{hyperref}
\usepackage{amsfonts}
\usepackage{amssymb}
\usepackage{mathrsfs}
\usepackage{subfigure}
\usepackage{ulem}

\begin{document}

\title{Phase-encoded measurement device independent quantum key distribution with practical spontaneous parametric-down-conversion sources}

\author{Chun Zhou} \affiliation{Zhengzhou Information Science and Technology Institute, Zhengzhou, 450004, China}

\author{Wan-Su Bao}\email[ ]{To whom correspondence should be addressed}
\email{2010thzz@sina.com}\affiliation{Zhengzhou Information Science and Technology Institute, Zhengzhou, 450004, China}

\author{Wei Chen}\email{kooky@mail.ustc.edu.cn}\affiliation{Key Laboratory of Quantum Information,University of Science and Technology of China, Hefei, 230026, China}

\author{Hong-Wei Li}\affiliation{Zhengzhou Information Science and Technology Institute, Zhengzhou, 450004, China}\affiliation{Key Laboratory of Quantum Information,University of Science and Technology of China, Hefei, 230026, China}

\author{Zhen-Qiang Yin}\affiliation{Key Laboratory of Quantum Information,University of Science and Technology of China, Hefei, 230026, China}

\author{Yang Wang}\affiliation{Zhengzhou Information Science and Technology Institute, Zhengzhou, 450004, China}

\author{Zheng-Fu Han}\email{zfhan@ustc.edu.cn}\affiliation{Key Laboratory of Quantum Information,University of Science and Technology of China, Hefei, 230026, China}

 \date{\today}

\begin{abstract}
Measurement-device-independent quantum key distribution (MDI-QKD) with weak coherent sources has been widely and meticulously analyzed. However, the analysis for MDI-QKD with spontaneous parametric-down-conversion sources (SPDCS) is incomplete. In this paper, we propose two passive decoy protocols suitable for parameter estimation in MDI-QKD using SPDCS. By accounting for practical parameters of SPDCS with thermal distribution, we present an investigation on the performances of MDI-QKD under the active three-intensity decoy protocol, the passive two-intensity decoy protocol and the modified passive three-intensity protocol respectively. Phase randomization, inherently prerequisite for decoy protocol, is taken into consideration for evaluating the overall quantum bit gain and quantum bit error rate. The numerical simulations show that the MDI-QKD using SPDCS with practical decoy protocols can be demonstrated comparable to the asymptotical case and has apparent superiority both in transmission distance and key generation rate compared to the one using weak coherent sources. Our results also indicate that the modified passive three-intensity decoy protocol can perform better than the active three-intensity decoy protocol in MDI-QKD using practical SPDCS.
\end{abstract}

\pacs{03.65.Ud, 03.65.Yz, 03.67.-a}
\maketitle

\section{introduction}
Quantum key distribution (QKD) is an art to make sure that the two legal communication parties can share the same key based on quantum mechanics. QKD has achieved great development both in theory and experiment \cite{Yamamoto,Stucki,Shuang,Jouguet,MaXS,Pan} since Bennett and Brassard presented the first QKD protocol, BB84 protocol \cite{BB84}. The ideal QKD has been theoretically proved to be unconditionally secure in different ways \cite{Lo,Shor,Renner,Inamori,GLLP,Tomamichel,Leverrier}, no matter how great the computation power and storage space the eavesdropper owns. However, practical QKD system undoubtedly exists sorts of imperfections. In fact, unconditional security of QKD never implies there is no restriction on the security in real situation. It holds under three important assumptions \cite{Scarani}, i.e., quantum mechanics are correct, authentication of classical communication is secure and all devices are believable. Some loopholes due to imperfections of setups will open windows for the Eve to acquire the secret key. Just because of this, some quantum hacking strategies have been subtly derived and successfully attacked the practical QKD system, such as time-shift attack \cite{Hacking1}, detector-blinding attack \cite{Hacking2}, wavelength-dependent attack \cite{Hacking3}, and so on \cite{Hacking4,Hacking5,Hacking6,Hacking7}.

Some countermeasures have been proposed to overcome these attacks. One approach is to derive efficient mathematical model to characterize the properties of devices and take imperfect factors into account in security proof as comprehensively as possible \cite{Scarani}. But from the angle of philosophy, it will be difficult to implement on a practical level. Another way is to employ the device-independent QKD (DI-QKD) \cite{DI1,DI2}, in which the security is based on the violation of a Bell inequality \cite{Clauser} without knowing the specifications of the devices used. However, it inherently requires a high detection efficiency to overcome the security loophole, which is hardly to achieve for practical single photon detectors, leading directly to a extremely low key rate at practical distances \cite{DI3}. Thus, people begins to look for feasible schemes for QKD that are intermediate between standard (device-dependent) QKD and DI-QKD \cite{DI4,DI5,MDI1,MDI2}.

Measurement-device independent QKD (MDI-QKD) scheme is recently proposed by Lo et al \cite{MDI1}, in which Alice and Bob both send photons to an untrusted third party who can even be an eavesdropper, say Eve. Eve performs a partial Bell-state measurement and announces the results to Alice and Bob for distilling a secret key. In MDI-QKD, the detection system can be considered as an oracle with input and output trusted, but no matter what has happened within it. Thus, MDI-QKD can remove all detector side channels, of which the security is guaranteed by the entanglement swapping techniques and reverse Einstein-Podolsky-Rosen (EPR) schemes \cite{MDI2}. Once MDI-QKD has been put forward, some modified schemes have been proposed \cite{MDI3,MDI5} and several experimental demonstrations have been performed \cite{MDI6,MDI7,MDI8,MDI9}.

Note the fact that the sources of MDI-QKD should be trusted, a complete characterization of the source is indispensable. The weak coherent source is commonly used as a replacement of the perfect single photon source. However, like the standard QKD, the MDI-QKD with weak coherent sources is also vulnerable to the photon-number-splitting attack due to the multi-photon fraction \cite{Brassard2000}. Fortunately, the decoy-state method \cite{decoy1, decoy2, decoy3} can be sufficiently applied to against this attack. More importantly, it can be used as a tool to efficiently estimate the contribution of the single-photon pulse. In Lo et al.'s seminal paper \cite{MDI1}, the ideal infinite decoy-state protocol is conducted to evaluate the contribution of the single-photon pulse. But their result is almost impossible to be realized in real situations because of the limited source. Later, some practical decoy-state protocols and their improvements have been proposed by many researchers combining such as finite-size effect, basis dependent imperfection or quantum memories \cite{MDI-d1,MDI-d2,Finite-MDI,MDI-d3,MDI-d4,MDI-d5}. But most of these results are based on weak coherent source (WCS) except for the three-intensity decoy-state method proposed by Wang \cite{MDI-d3}. For another candidate of photon sources within reach of current technology, i.e., the conditional generation of single photons based on parametric down-conversion \cite{SPDC1,SPDC2,SPDC3}, the performance of decoy-state MDI-QKD remains an issue of common concern.

Recently, we notice that Wang et al. \cite{MDI-d6} derived a formula for estimating the single-photon contribution for the MDI-QKD with Poisson distributed heralded single photon source (HSPS). However, in this paper, we will analyze the case when the spontaneous parametric-down-conversion sources (SPDCS) under a practical photon number distribution (PND) are used in MDI-QKD. Here, the PND is determined by a SPDCS with a threshold detector. Just simply considering the PND when threshold detector is triggered, we can apply the active three-intensity decoy-state protocol to estimate the single-photon's contribution. While also taking the non-triggered PND into account, the passive decoy state protocol originally proposed by Adachi et al. is conducted \cite{Adachi}. Note that full phase randomization of each individual pulse is a crucial assumption in security proofs of decoy state QKD and active phase randomization is implemented to protect against attacks on imperfect sources in recent experimental demonstration of MDI-QKD\cite{MDI9}. Thus, differing from the analysis of Wang et al. \cite{MDI-d6}, our evaluations for the overall gain and quantum bit error rate (QBER) are based on full phase randomization, which is internally demanded for decoy-state method. So our formulas for the overall gain and QBER are more stringent.

In this paper, inspired from the passive decoy-state method presented by Adachi et al.\cite{Adachi}, we raise two passive decoy protocols, the passive two-intensity decoy protocol and the modified passive three-intensity decoy protocol, which can be applied in MDI-QKD using SPDCS with threshold detectors. Taking the phase-encoded scheme proposed by Ma et al. for an example\cite{MDI5}, we analyze the performances of MDI-QKD with practical SPDCS when different decoy protocols are conducted. Numerical simulations show that the MDI-QKD using SPDCS has apparent superiority both in transmission distance and key generation rate compared to the one using weak coherent sources. Moreover, both of the modified passive three-intensity decoy protocol and active three-intensity decoy protocol\cite{MDI-d3} can be demonstrated comparable to the asymptotical case when active infinite decoy states are used. Our results also indicate the modified passive three-intensity decoy protocol can perform better than the active three-intensity decoy protocol in MDI-QKD using SPDCS with a practical photon number distribution. The rest of paper is organized as follows. In Sec. II, we shall briefly review the SPDCS with practical photon number distribution. The formulas for calculating the single-photon yield and error rate using active three-intensity decoy protocol are introduced in Sec. III. We derive the formulas for the passive two-intensity decoy protocol and modified passive three-intensity decoy protocol respectively in Sec. IV and Sec. V. We perform numerical simulations in Sec. VI, followed by a statistical fluctuation analysis in Sec. VII. We conclude the paper in Sec. VIII.

\section{practical spontaneous parametric-down-conversion source}
As to our knowledge, spontaneous parametric-down-conversion (SPDC) processes are widely used as the sources in QKD such as the triggered or heralded single-photon source\cite{Adachi,Horikiri}, and the entanglement source\cite{Etangle,Erven}. Up to now, there exist some promising implementation demonstrations of QKD  based on SPDC processes\cite{Etangle,Trigger1,Erven}. In this paper, as is shown in Fig.1, we consider the case when SPDC processes are used as triggered single-photon sources in phase-encoded MDI-QKD scheme proposed by Ma et al.\cite{MDI5}(other experimental schemes like the original polarization-based protocol\cite{MDI1} and the phase-encoded protocol proposed by Tamaki et.al\cite{MDI3} are also feasible). We remark that the security of scheme shown in Fig.1 relies on detectors $D_A$ and $D_B$ to be trusted, which maybe introduce side channel attack if the laboratories are untrusted\cite{Passive-2}. However, this type of attacks can be avoided if Alice and Bob have secret area respectively.
\begin{figure}[!h]\center
\resizebox{8.5cm}{!}{
\includegraphics{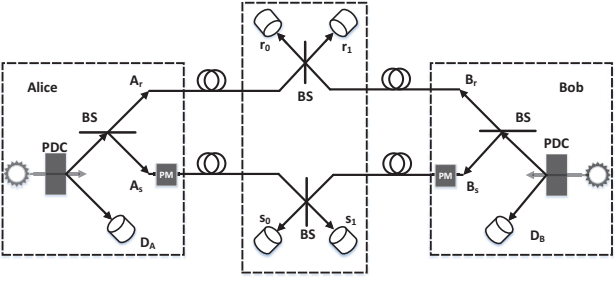}}
\caption{The schematic diagram of the phase-encoded MDI-QKD protocol using triggered single-photon sources based on SPDC processes. Here, PDC represents the non-degenerate SPDC process, BS stands for 50:50 beam splitter and PM stands for the phase modulator. $D_A$ and $D_B$ are the threshold detectors used for triggering signal pulses, $r_0$, $r_1$, $s_0$ and $s_1$ are the four single-photon detectors for performing Bell state measurement.}
\end{figure}

 The two identical sources with sub-Poissonian distribution are emitted from non-degenerate SPDC process\cite{SPDC1}. This type of SPDC processes creates the two-mode state
\begin{equation}
\begin{array}{lll}
(\cosh\chi)^{-1}\sum\limits_{n=0}^{\infty }{{{(\tanh \chi )}^{n}}{{e}^{in\theta }}\left| n,n \right\rangle }
\end{array}
\end{equation}
Suppose the intensity $\mu$ of the source to be $\sinh^2\chi$, then the above description simplifies to
\begin{equation}
\begin{array}{lll}
\sum\limits_{n=0}^{\infty}{\sqrt{\frac{{{\mu}^{n}}}{{{(1+\mu )}^{n+1}}}}{{e}^{in\theta }}\left|n,n\right\rangle}
\end{array}
\end{equation}
When Alice(Bob) monitors one mode of her(his) SPDC source with a practical threshold detector described by detection efficiency $\eta_A$($\eta_B$) and dark count rate $d_A$($d_B$), Alice's(Bob's) other mode changes to a mixed state under the condition that the detector has been successfully triggered. This state can be described as\cite{Trigger1,HSPS2}
\begin{equation}
\label{HSPS-A}
\begin{array}{lll}
{{\left|\psi(\mu_A,\theta_A)\ \right\rangle }_{A}}=\sqrt{\frac{P_{\mu_A}^{cor}}{1+\mu_A}\cdot \frac{d_A}{P_{\mu_A}^{post}}+(1-P_{\mu_A}^{cor})}\left|0\right\rangle \\+\sum\limits_{n=1}^{\infty }{\sqrt{P_{\mu_A}^{cor}\cdot\frac{{{\mu_A}^{n}}}{{{(1+\mu_A )}^{n+1}}}\cdot\frac{1-{{(1-{{\eta}_{A}})}^{n}}+{{d}_{A}}}{P_{\mu_A}^{post}}}{{e}^{in\theta_A}}\left| n \right\rangle},
\end{array}
\end{equation}
\begin{equation}
\label{HSPS-B}
\begin{array}{lll}
{{\left|\psi(\mu_B,\theta_B)\ \right\rangle }_{B}}=\sqrt{\frac{P_{\mu_B}^{cor}}{1+\mu_B}\cdot \frac{d_B}{P_{\mu_B}^{post}}+(1-P_{\mu_B}^{cor})}\left|0\right\rangle \\+\sum\limits_{n=1}^{\infty }{\sqrt{P_{\mu_B}^{cor}\cdot\frac{{{\mu_B}^{n}}}{{{(1+\mu_B )}^{n+1}}}\cdot\frac{1-{{(1-{{\eta}_{B}})}^{n}}+{{d}_{B}}}{P_{\mu_B}^{post}}}{{e}^{in\theta_B}}\left| n \right\rangle}
\end{array}
\end{equation}
where $\mu_A$($\mu_B$) is the mean photon number of Alice's (Bob's) one mode before
triggering, $P_{\mu_A}^{post}=1+d_A-\frac{1}{1+\mu_A\eta_A}$ ($P_{\mu_B}^{post}=1+d_B-\frac{1}{1+\mu_B\eta_B}$) denotes the post-selection probability, and $P_{\mu_A}^{cor}$($P_{\mu_B}^{cor}$) is the correlation rate of photon pairs for Alice's(Bob's) source with intensity of $\mu_A$($\mu_B$), i.e. the probability that one can predict the existence of a heralded photon given a triggered one. After phase randomization(integrating $\theta$ over $[0,2\pi]$), Alice and Bob's heralded mixed state shown as Eq.(\ref{HSPS-A})and Eq.(\ref{HSPS-B}) can be described by the following photon number distribution
\begin{eqnarray}
\label{HSPS-C}
P_n(\mu_x)=\left\{\begin{array}{cc}
1-P_{\mu_x}^{cor}+\frac{d_xP_{\mu_x}^{cor}}{(1+\mu_x)P_{\mu_x}^{post}} & for\ n=0 \\
\frac{{\mu_x^n}[1-(1-\eta_x)^n+d_x]P_{\mu_x}^{cor}}{(1+\mu_x)^{n+1}P_{\mu_x}^{post}} & for\ n\ge 1 \\
\end{array}\right.
\end{eqnarray}
where $x=A$($x=B$) denotes the case of triggered state output from Alice(Bob).

\section{active three-intensity decoy-state protocol}
Once the two-pulse state from Alice and Bob's SPDCS arrives at the relay(controlled by an untrusted eavesdropper Eve)for detection, Eve will announce whether the two-pulse state has caused a successful Bell state measurement according to the detection results. Those states corresponding to successful measurement events at the relay will be post selected as the sifted keys and be input to the postprocessing procedure. The most key point is how to estimate the single-photon contributions and errors based on practical experiments. Luckily, the three-intensity decoy-state protocol for MDI-QKD recently proposed by Wang\cite{MDI-d3} can be introduced for this purpose. In this protocol, Alice(Bob) need to actively and randomly modulate her(his) sources to prepare different states with intensity of 0,$\mu_A$ and $\mu_A'$(0,$\mu_B$ and $\mu_B'$). Here, $\mu_A'$($\mu_B'$) is the intensity of signal state, 0 is the intensity of vacuum states,  $\mu_A$($\mu_B$)is the intensity of decoy state and $\mu_A'>\mu_A$($\mu_B'>\mu_B$). According to Wang's theory, if $\frac{P_1(\mu'_A)P_2(\mu'_B)}{P_1(\mu_A)P_2(\mu_B)}\le \frac{P_2(\mu'_A)P_1(\mu'_B)}{P_2(\mu_A)P_1(\mu_B)}$ holds true, the low bound of successful single-photon counting rate in the $Z$ basis and upper bound of single-photon error rate in the $X$ basis can be estimated by\cite{MDI-d3}
\begin{equation}
\label{active-1}
\begin{array}{lll}
Y^Z_{11}\geq \frac{P_1(\mu'_A)P_2(\mu'_B) (Q_{\mu_A\mu_B}-\overline{Q}_0)-P_1(\mu_A)P_2(\mu_B) (Q_{\mu'_A\mu'_B}-\overline{Q}'_0)}{P_1(\mu'_A)P_1(\mu_A)[P_2(\mu'_B)P_1(\mu_B)-P_2(\mu_B)P_1(\mu'_B)]}
\end{array}
\end{equation}

\begin{equation}
\label{active-2}
\begin{array}{lll}
E^X_{11}\leq \frac{E^X_{\mu_A\mu_B}Q^X_{\mu_A\mu_B}-P_0(\mu_A)E^X_{0\mu_B}Q^X_{0\mu_B}-P_0(\mu_B)E^X_{\mu_A0}Q^X_{\mu_A0}}{P_1(\mu_A)P_1(\mu_B)Y^X_{11}}\\
\quad\quad\quad+\frac{P_0(\mu_A)P_0(\mu_B)E_{00}Y_{00}}{P_1(\mu_A)P_1(\mu_B)Y^X_{11}}
\end{array}
\end{equation}
Here, $Q_{\mu_A\mu_B}$($Q_{\mu'_A\mu'_B}$) denotes the overall gain when the intensity of Alice's source is $\mu_A$($\mu'_A$) and that of Bob's source is $\mu_B$($\mu'_B$), $\overline{Q}_0=P_0(\mu_A)Q_{0\mu_B}+P_0(\mu_B)Q_{\mu_A 0}-P_0(\mu_A)P_0(\mu_B)Y_{00}$, $\overline{Q}'_0=P_0(\mu'_A)Q_{0\mu'_B}+P_0(\mu'_B)Q_{\mu'_A 0}-P_0(\mu'_A)P_0(\mu'_B)Y_{00}$, $Q_{0\mu_B}$, $Q_{\mu_A 0}$, $Q_{0\mu'_B}$ and $Q_{\mu'_A 0}$ are the overall gain under different sources where $0$ represents that the intensity of the corresponding source is vacuum; $Y_{00}$ is the yield of a background noise; $Q^{X}_{\mu_A\mu_B}$, $Q^{X}_{0\mu_B}$ and $Q^{X}_{\mu_A0}$ is the overall gain in the $X$ basis under different sources ; $E^{X}_{\mu_A\mu_B}$, $E^{X}_{0\mu_B}$ and $E^{X}_{\mu_A0}$ is the overall quantum bit error rate of events in the $X$ basis under different sources ; $E_{00}=1/2$ is the error rate of a random noise.

Thus, the key rate for the active three intensity decoy-state protocol can be given by\cite{MDI-d3}
\begin{equation}
\label{active-3}
\begin{array}{lll}
R=P_1(\mu'_A)P_1(\mu'_B)Y^Z_{11}(1-H(E^X_{11}))\\
\quad\quad-Q^Z_{\mu'_A\mu'_B}f(E^Z_{\mu'_A\mu'_B})H(E^Z_{\mu'_A\mu'_B}).
\end{array}
\end{equation}
where $H(x)=-xlog_2(x)-(1-x)log_2(1-x)$ denotes the binary Shannon entropy function, $Q^Z_{\mu'_A\mu'_B}$ and $E^Z_{\mu'_A\mu'_B}$ are the overall gain and error rate in the $Z$ basis under source of $\mu'_A\mu'_B$ . $f(x)$ is the error correction efficiency.

\section{passive two-intensity decoy-state protocol}
Active decoy-state protocol typically needs a variable optical attenuator (VOA) in signal sender's side to independently and randomly vary the intensity of each signal state. However, in some scenarios, if the VOA is not correctly designed, it may occur that  some physical parameters of the sending pulses rely on the particular setting selected \cite{Passive-4}, which could bring threat to the security of the active schemes. Therefore passive preparation of intensity might be practically desirable to some extent \cite{Passive-1}. In \cite{Adachi}, Adachi et al. present an efficient passive decoy-state proposal for the QKD based on SPDCS, where only a simple threshold detector is used. Later, Ma and Lo \cite{Passive-2} combined the results of \cite{Adachi} and \cite{Passive-1} to the most common case. In 2009, Curty et al. \cite{Passive-3} generalized the passive decoy idea and proposed a new scheme for the case when weak coherent source is used. Here, inspiring from the idea proposed by Adachi et al. \cite{Adachi}, we shall apply the passive decoy protocol to the MDI-QKD using SPDCS. The experimental setup, as is shown in Fig.1, is kept unchanged as the active three intensity decoy-state protocol, except keeping the detectors in the relay to work no matter whether there is a trigger signal or not in Alice and Bob's threshold detectors. Therefore, the non-triggered events, acting as the role of decoy states, can be used to estimate the single-photon contribution. In this case, the signal $n$-photon events can be divided into two parts, the triggered events with probability of $P^T_n(\mu_x)$ and the non-triggered events with probability of $P_n^{NT}(\mu_x)$
\begin{eqnarray}
\label{Passive-01}
{P_{n}^{T}}({{\mu}_{x}})=\left\{\begin{array}{cc}
\frac{1-P_{\mu_x}^{cor}}{2}+\frac{P_{\mu_x}^{cor}d_x}{(1+\mu_x)} & for\ n=0 \\
\frac{{\mu_x^n}[1-(1-\eta_x)^n+d_x]P_{\mu_x}^{cor}}{(1+\mu_x)^{n+1}} & for\ n\ge 1 \\
\end{array}\right.
\end{eqnarray}
\begin{eqnarray}
\label{Passive-02}
{P_{n}^{NT}}({{\mu}_{x}})=\left\{\begin{array}{cc}
\frac{1-P_{\mu_x}^{cor}}{2}+\frac{P_{\mu_x}^{cor}(1-d_x)}{(1+\mu_x)} & for\ n=0 \\
\frac{{\mu_x^n}[(1-\eta_x)^n-d_x]P_{\mu_x}^{cor}}{(1+\mu_x)^{n+1}} & for\ n\ge 1 \\
\end{array}\right.
\end{eqnarray}
where $x=A$ or $x=B$.
If we consider the above photon number distribution shown in Eq.(\ref{Passive-01}) and Eq.(\ref{Passive-02}), the corresponding overall counting rates of triggered events and non-triggered events can be expressed as
\begin{equation}
\label{Passive-1}
\begin{array}{lll}
Q_{\mu_A\mu_B}^{(t)}=\sum\limits_{n=0,m=0}^{\infty}{{Q^{(t)}_{nm}}}\\
\quad\quad\quad=Q_{\mu_A0}^{(t)}+Q_{0\mu_B}^{(t)}-Q_{00}^{(t)}+Q_{11}^{(t)}\\
\quad\quad\quad\quad+\sum\limits_{m=2}^{\infty}{{Q^{(t)}_{1m}}}+\sum\limits_{n=2}^{\infty}{{Q^{(t)}_{n1}}}+\sum\limits_{n=2,m=2}^{\infty}{{Q^{(t)}_{nm}}}
\end{array}
\end{equation}
\begin{equation}
\label{Passive-2}
\begin{array}{lll}
Q_{\mu_A\mu_B}^{(nt)}=\sum\limits_{n=0,m=0}^{\infty}{{Q^{(nt)}_{nm}}}\\
\quad\quad\quad=Q_{\mu_A0}^{(nt)}+Q_{0\mu_B}^{(nt)}-Q_{00}^{(nt)}+Q_{11}^{(nt)}\\
\quad\quad\quad\quad+\sum\limits_{m=2}^{\infty}{{Q^{(nt)}_{1m}}}+\sum\limits_{n=2}^{\infty}{{Q^{(nt)}_{n1}}}+\sum\limits_{n=2,m=2}^{\infty}{{Q^{(nt)}_{nm}}}
\end{array}
\end{equation}
where $Q^{(t)}_{nm}=Y_{nm}P^{T}_n(\mu_A)P^{T}_m(\mu_B)$ is the triggered gain when Alice sends $n$-photon pulse and Bob sends $m$-photon pulse, $Q^{(nt)}_{nm}=Y_{nm}P^{NT}_n(\mu_A)P^{NT}_m(\mu_B)$ is the non-triggered gain when Alice sends $n$-photon pulse and Bob sends $m$-photon pulse, $Q_{\mu_A0}^{(t)}$ is the triggered overall gain when the intensity of Alice's source is $\mu_A$ and that of Bob's source is $0$. The meaning of $Q_{0\mu_B}^{(t)}$, $Q_{\mu_A0}^{(nt)}$ and $Q_{0\mu_B}^{(nt)}$ is similar to $Q_{\mu_A0}^{(t)}$, $Q_{00}^{(t)}=P^{T}_0(\mu_A)P^{T}_0(\mu_B)Y_{00}$ and $Q_{00}^{(nt)}=P^{NT}_0(\mu_A)P^{NT}_0(\mu_B)Y_{00}$ are the gains from background noises. Define $r_{nm}=r_nr_m$ where $r_n=\frac{P^T_n(\mu_A)}{P^{NT}_n(\mu_A)}$ and $r_m=\frac{P^T_m(\mu_B)}{P^{NT}_m(\mu_B)}$, then $Q^{(t)}_{nm}=r_{nm}Q^{(nt)}_{nm}$. From a mathematical analysis, we can obtain that $0<r_0<r_1<r_2<r_3<\cdots$. Thus, it is obvious that $r_{nm}<min\{r_{n(m+1)},r_{(n+1)m}\}\leq max\{r_{n(m+1)},r_{(n+1)m}\}<r_{(n+1)(m+1)}$ for $n\geq1, m\geq1$. Define $r_{min}=min(r_{12},r_{21})$, then $r_{min}Q^{(nt)}_{nm}\leq Q^{(t)}_{nm}$ for $n,m \geq 1$. Applying Eq.(\ref{Passive-1}) and Eq.(\ref{Passive-2}) leads to $r_{min}[Q_{\mu_A\mu_B}^{(nt)}-Q_{\mu_A0}^{(nt)}-Q_{0\mu_B}^{(nt)}+Q_{00}^{(nt)}-Q_{11}^{(nt)}]\leq Q_{\mu_A\mu_B}^{(t)}-Q_{\mu_A0}^{(t)}-Q_{0\mu_B}^{(t)}+r_{00}Q_{00}^{(nt)}-r_{11}Q_{11}^{(nt)}$, we thus obtain the minimum value of $Q^{(nt)}_{11}$ as a function of the parameter $\alpha\equiv\frac{Q_{00}^{(nt)}}{Q_{\mu_A\mu_B}^{(nt)}}$:
\begin{equation}
\label{Passive-3}
\begin{array}{lll}
\frac{Q_{11}^{(nt)}}{Q_{\mu_A\mu_B}^{(nt)}}\geq \frac{(r_{min}-r_{00})\alpha+r_{min}Q^{(nt)}_{\Delta}-Q^{(t)}_{\Delta}}{r_{min}-r_{11}}\triangleq \xi(\alpha),
\end{array}
\end{equation}
where $Q^{(nt)}_{\Delta}=\frac{Q_{\mu_A\mu_B}^{(nt)}-Q_{\mu_A0}^{(nt)}-Q_{0\mu_B}^{(nt)}}{Q_{\mu_A\mu_B}^{(nt)}}$ and $Q^{(t)}_{\Delta}=\frac{Q_{\mu_A\mu_B}^{(t)}-Q_{\mu_A0}^{(t)}-Q_{0\mu_B}^{(t)}}{Q_{\mu_A\mu_B}^{(nt)}}$.

Note that the overall quantum bit error rates of triggered and non-triggered events are calculated by $E_{\mu_A\mu_B}^{(t)}=\sum\limits_{n=0,m=0}^{\infty}{\frac {Q^{(t)}_{nm}E_{nm}}{Q_{\mu_A\mu_B}^{(t)}}}$ and $E_{\mu_A\mu_B}^{(nt)}=\sum\limits_{n=0,m=0}^{\infty}{\frac {Q^{(nt)}_{nm}E_{nm}}{Q_{\mu_A\mu_B}^{(nt)}}}$, we can also get the upper bound of $e_{11}$ as
\begin{equation}
\label{Passive-4}
\begin{array}{lll}
e_{11}\leq \frac{E^{(t)}_{\Delta}+\alpha r_{00}E_{00}}{r_{11}\xi(\alpha)}\triangleq\epsilon_{t}(\alpha)
\end{array}
\end{equation}
where $E^{(t)}_{\Delta}=\frac{Q_{\mu_A\mu_B}^{(t)}E_{\mu_A\mu_B}^{(t)}-Q_{\mu_A0}^{(t)}E_{\mu_A0}^{(t)}-Q_{0\mu_B}^{(t)}E_{0\mu_B}^{(t)}}{Q_{\mu_A\mu_B}^{(nt)}}$ and $E_{00}=1/2$ is the error rate of a random noise.

In a similar way, we have another bound
\begin{equation}
\label{Passive-5}
\begin{array}{lll}
e_{11}\leq \frac{E^{(nt)}_{\Delta}+\alpha E_{00}}{\xi(\alpha)}\triangleq\epsilon_{nt}(\alpha)
\end{array}
\end{equation}
where $E^{(nt)}_{\Delta}=\frac{Q_{\mu_A\mu_B}^{(nt)}E_{\mu_A\mu_B}^{(nt)}-Q_{\mu_A0}^{(nt)}E_{\mu_A0}^{(nt)}-Q_{0\mu_B}^{(nt)}E_{0\mu_B}^{(nt)}}{Q_{\mu_A\mu_B}^{(nt)}}$.

Combining these two upper bounds, we have
\begin{equation}
\label{Passive-6}
\begin{array}{lll}
e_{11}\leq min\{\epsilon_{t}(\alpha), \epsilon_{nt}(\alpha)\}\triangleq \epsilon(\alpha)
\end{array}
\end{equation}
where $0\leq\alpha\leq min\{\frac{2Q_{\mu_A\mu_B}^{(t)}E_{\mu_A\mu_B}^{(t)}}{r_{00}Q_{\mu_A\mu_B}^{(nt)}}, 2E_{\mu_A\mu_B}^{(nt)}\}$.

Consequently, the key rate from the triggered events can be given by
\begin{equation}
\label{Passive-7}
\begin{array}{lll}
R^{(t)}=Q_{\mu_A\mu_B}^{(nt)}\min\limits_{\alpha=0}\{r_{00}\alpha+r_{11}\xi(\alpha)[1-H(\epsilon(\alpha))]\}\\
\quad\quad\quad-Q_{\mu_A\mu_B}^{(t)}f(E_{\mu_A\mu_B}^{(t)})H(E_{\mu_A\mu_B}^{(t)}),
\end{array}
\end{equation}
and the key rate from both the triggered and non-triggered events is given by
\begin{equation}
\label{Passive-8}
\begin{array}{lll}
R^{(both)}\\
=Q_{\mu_A\mu_B}^{(nt)}\min\limits_{\alpha=0}\{(1+r_{00})\alpha+(1+r_{11})\xi(\alpha)[1-H(\epsilon(\alpha))]\}\\
\quad-Q_{\mu_A\mu_B}^{(t)}f(E_{\mu_A\mu_B}^{(t)})H(E_{\mu_A\mu_B}^{(t)})\\
\quad-Q_{\mu_A\mu_B}^{(nt)}f(E_{\mu_A\mu_B}^{(nt)})H(E_{\mu_A\mu_B}^{(nt)}).
\end{array}
\end{equation}

Therefore, the final key rate of the passive one intensity decoy-state protocol is given by
\begin{equation}
\label{Passive-9}
\begin{array}{lll}
R=\max\{R^{(t)},R^{(both)}\}.
\end{array}
\end{equation}

\section{modified passive three-intensity decoy-state protocol}
For the active three-intensity decoy-state protocol, the basic assumption is that the yield of decoy states is equal to that of signal states, i.e., $Y_{nm}=Y'_{nm}$ for $n\geq1, m\geq1$. And for the passive two-intensity decoy-state protocol, the assumption is that the yield of triggered states is equal to that of non-triggered states, i.e., $Y_{nm}^{T}=Y_{nm}^{NT}$ for $n\geq1, m\geq1$. Similarly, considering the case when active decoy states are used in passive decoy protocol, we can also assume that the yield of triggered decoy states is equal to that of non-triggered signal states, i.e., $Y_{nm}^{T}={Y'}_{nm}^{NT}$ for $n\geq1, m\geq1$. Based on this assumption, we improve the passive two-intensity decoy-state protocol and propose a new protocol called modified passive three-intensity decoy-state protocol.

In our new protocol, Alice(Bob) also need to actively and randomly modulate her(his) sources to prepare different states with intensity of 0,$\mu_A$ and $\mu_A'$(0,$\mu_B$ and $\mu_B'$). Here, $\mu_A'$($\mu_B'$) is the intensity of signal states, 0 is the intensity of vacuum states,  $\mu_A$($\mu_B$)is the intensity of decoy states and $\mu_A'>\mu_A$($\mu_B'>\mu_B$). The set of signal states is divided into two parts, one part is the set of triggered signal states and the other is the set of non-triggered signal states. And the set of decoy states is also divided into the part of triggered decoy states and the part of non-triggered decoy states. For convenience, we choose the non-triggered signal states and triggered decoy states to derive the yield of single photons. And we require that
\begin{equation}
\label{M-Passive-1}
\begin{array}{lll}
\frac{{P_{k}^{T}}(\mu_A')}{{P_{k}^{T}}(\mu_A)}\geq\frac{{P_{2}^{T}}(\mu_A')}{{P_{2}^{T}}(\mu_A)}\geq\frac{{P_{1}^{T}}(\mu_A')}{{P_{1}^{T}}(\mu_A)},\\\frac{{P_{k}^{NT}}(\mu_A')}{{P_{k}^{NT}}(\mu_A)}\geq\frac{{P_{2}^{NT}}(\mu_A')}{{P_{2}^{NT}}(\mu_A)}\geq\frac{{P_{1}^{NT}}(\mu_A')}{{P_{1}^{NT}}(\mu_A)},\\
\frac{{P_{k}^{T}}(\mu_B')}{{P_{k}^{T}}(\mu_B)}\geq\frac{{P_{2}^{T}}(\mu_B')}{{P_{2}^{T}}(\mu_B)}\geq\frac{{P_{1}^{T}}(\mu_B')}{{P_{1}^{T}}(\mu_B)},\\\frac{{P_{k}^{NT}}(\mu_B')}{{P_{k}^{NT}}(\mu_B)}\geq\frac{{P_{2}^{NT}}(\mu_B')}{{P_{2}^{NT}}(\mu_B)}\geq\frac{{P_{1}^{NT}}(\mu_B')}{{P_{1}^{NT}}(\mu_B)}. \end{array}
\end{equation}
The convex forms of the overall triggered decoy events and non-triggered signal events can be expressed as
\begin{equation}
\label{M-Passive-2}
\begin{array}{lll}
\Omega_{\mu_A\mu_B}^{(t)}=\tilde{\Omega}_0^{T}+{P_{1}^{T}}(\mu_A){P_{1}^{T}}(\mu_B)\rho_1 \otimes \rho_1\\
\quad\quad\quad\quad+P_{1}^{T}(\mu_A)\rho_1 \otimes(\sum\limits_{n=2}^{\infty}{P_{n}^{T}(\mu_B)|n\rangle\langle n|})\\
\quad\quad\quad\quad+P_{1}^{T}(\mu_B)\rho_1 \otimes(\sum\limits_{m=2}^{\infty}{P_{m}^{T}(\mu_A)|m\rangle\langle m|})\\
\quad\quad\quad\quad+\sum\limits_{n=2,m=2}^{\infty}{P_{n}^{T}(\mu_B)P_{m}^{T}(\mu_A)|n\rangle\langle n| \otimes |m\rangle\langle m|}
\end{array}
\end{equation}
\begin{equation}
\label{M-Passive-3}
\begin{array}{lll}
\Omega_{\mu_A'\mu_B'}^{(nt)}=\tilde{\Omega}_0^{NT}+{P_{1}^{NT}}(\mu_A'){P_{1}^{NT}}(\mu_B')\rho_1 \otimes \rho_1\\
\quad\quad\quad\quad+P_{1}^{NT}(\mu_A')\sum\limits_{n=2}^{\infty}{P_{n}^{NT}(\mu_B')|n\rangle\langle n|}\\
\quad\quad\quad\quad+P_{1}^{NT}(\mu_B')\sum\limits_{m=2}^{\infty}{P_{m}^{NT}(\mu_A')|m\rangle\langle m|}\\
\quad\quad\quad\quad+\sum\limits_{n=2,m=2}^{\infty}{P_{n}^{NT}(\mu_B')P_{m}^{NT}(\mu_A')|n\rangle\langle n| \otimes |m\rangle\langle m|}
\end{array}
\end{equation}
where $\tilde{\Omega}_0^{T}={P_{0}^{T}}(\mu_B)\Omega_{\mu_A0}^{T}+{P_{0}^{T}}(\mu_A)\Omega_{0\mu_B}^{T}-{P_{0}^{T}}(\mu_A){P_{0}^{T}}(\mu_B)\Omega_{00}^{T}$, $\tilde{\Omega}_0^{NT}={P_{0}^{NT}}(\mu_B')\Omega_{\mu_A'0}^{NT}+{P_{0}^{NT}}(\mu_A')\Omega_{0\mu_B'}^{NT}-{P_{0}^{NT}}(\mu_A'){P_{0}^{NT}}(\mu_B')\Omega_{00}^{NT}$, $\Omega_{\mu_A0}^{T}$ and $\Omega_{\mu_A0}^{NT}$ denote the triggered and non-triggered two-pulse state respectively when Alice sends pulse with intensity of $\mu_A$ and Bob sends vacuum pulse, $\Omega_{0\mu_B}^{T}$ and $\Omega_{0\mu_B}^{NT}$ denote the triggered and non-triggered two-pulse state respectively when Alice sends vacuum pulse and Bob sends pulse with intensity of $\mu_A$,
$\Omega_{00}^{T}$ and $\Omega_{00}^{NT}$ denote the triggered and non-triggered two-pulse state respectively when Alice and Bob both send vacuum pulse. Here, the state $\rho_1 \otimes \rho_1$ leads to the yield $Y_{11}$. Applying the same analysis as Wang's theory\cite{MDI-d3}, we can also derive a bound on $Y_{11}$ based on Eq.(\ref{M-Passive-2}) and Eq.(\ref{M-Passive-3}). Under the condition that $\frac{P_1^{NT}(\mu'_A)P_2^{NT}(\mu'_B)}{P_1^{T}(\mu_A)P_2^{T}(\mu_B)}\le\frac{P_2^{NT}(\mu'_A)P_1^{NT}(\mu'_B)}{P_2^{T}(\mu_A)P_1^{T}(\mu_B)}$, $Y^Z_{11}$ and $E^X_{11}$ can be estimated by
\begin{equation}
\label{M-Passive-4}
\begin{array}{lll}
Y^Z_{11}\geq \frac{P_1^{T}(\mu_A)P_2^{T}(\mu_B)(Q_{\mu'_A\mu'_B}^{NT}-\tilde{Q}_0^{NT})}{P_1^{NT}(\mu'_A)P_1^{T}(\mu_A)[P_2^{T}(\mu_B)P_1^{NT}(\mu'_B)-P_2^{NT}(\mu'_B)P_1^{T}(\mu_B)]}\\
\quad\quad\quad-\frac{P_1^{NT}(\mu'_A)P_2^{NT}(\mu'_B)(Q_{\mu_A\mu_B}^{T}-\tilde{Q}_0^{T})}{P_1^{NT}(\mu'_A)P_1^{T}(\mu_A)[P_2^{T}(\mu_B)P_1^{NT}(\mu'_B)-P_2^{NT}(\mu'_B)P_1^{T}(\mu_B)]},
\end{array}
\end{equation}

\begin{equation}
\label{M-Passive-5}
\begin{array}{lll}
E^X_{11}\leq \frac{E^T_{\mu_A\mu_B}Q^T_{\mu_A\mu_B}-P_0^T(\mu_A)E^T_{0\mu_B}Q^T_{0\mu_B}-P_0^T(\mu_B)E^T_{\mu_A0}Q^T_{\mu_A0}}{P_1^T(\mu_A)P_1^T(\mu_B)Y_{11}}\\
\quad\quad\quad+\frac{P_0^T(\mu_A)P_0^T(\mu_B)E_{00}Y_{00}}{P_1^T(\mu_A)P_1^T(\mu_B)Y_{11}},
\end{array}
\end{equation}
where $\tilde{Q}_0^{T}={P_{0}^{T}}(\mu_B)Q_{\mu_A0}^{T}+{P_{0}^{T}}(\mu_A)Q_{0\mu_B}^{T}-Q_{00}^{T}$, $\tilde{Q}_0^{NT}={P_{0}^{NT}}(\mu_B')Q_{\mu_A'0}^{NT}+{P_{0}^{NT}}(\mu_A')Q_{0\mu_B'}^{NT}-Q_{00}^{NT}$, $Q_{\mu_A\mu_B}^{T}$ is the triggered gain when Alice sends pulses with intensity of $\mu_A$ and Bob sends pulses with intensity of $\mu_B$, $Q_{\mu_A'\mu_B'}^{NT}$ is the non-triggered gain when Alice sends pulses with intensity of $\mu_A'$ and Bob sends pulses with intensity of $\mu_B'$, $Q_{\mu_A0}^T$ is the triggered overall gain when the intensity of Alice's source is $\mu_A$ and that of Bob's source is vacuum. The meaning of $Q_{0\mu_B}^{T}$, $Q_{\mu_A'0}^{NT}$ and $Q_{0\mu_B'}^{NT}$ is similar to $Q_{\mu_A0}^{T}$. $E^T_{\mu_A\mu_B}$ is the overall quantum bit error rate of triggered events in the $X$ basis when Alice sends pulses with intensity of $\mu_A$ and Bob sends pulses with intensity of $\mu_B$. $E^T_{0\mu_B}$, $E^T_{\mu_A0}$ and $E_{00}$ is the overall quantum bit error rate of triggered events in the $X$ basis when Alice and Bob send pulses with different intensities. $Q_{00}^{T}=P^{T}_0(\mu_A)P^{T}_0(\mu_B)Y_{00}$ and $Q_{00}^{NT}=P^{NT}_0(\mu_A')P^{NT}_0(\mu_B')Y_{00}$ are the gains from background noises.

Consequently, the key rate from the triggered events of signal pulses can be given by
\begin{equation}
\label{M-Passive-6}
\begin{array}{lll}
R^{T}=\tilde{Q}_0^{T}+P_1^{T}(\mu'_A)P_1^{T}(\mu'_B)Y^Z_{11}[1-H(E^X_{11})]\\
\quad\quad\quad-Q^{Z,T}_{\mu'_A\mu'_B}f(E^{Z,T}_{\mu'_A\mu'_B})H(E^{Z,T}_{\mu'_A\mu'_B}),
\end{array}
\end{equation}
where $Y^Z_{11}$ is the single photon yield in the $Z$ basis, $Q^{Z,T}_{\mu'_A\mu'_B}$ and $E^{Z,T}_{\mu'_A\mu'_B}$ is the overall triggered gain and quantum bit error rate in the $Z$ basis. The key rate from both the triggered and non-triggered events of signal pulses is given by
\begin{equation}
\label{M-Passive-7}
\begin{array}{lll}
R^{Both}=\tilde{Q}_0^{T}+\tilde{Q}_0^{NT}+P_1^{T}(\mu'_A)P_1^{T}(\mu'_B)Y^Z_{11}[1-H(E^X_{11})]\\
\quad\quad\quad\quad+P_1^{NT}(\mu'_A)P_1^{NT}(\mu'_B)Y^Z_{11}[1-H(E^X_{11})]\\
\quad\quad\quad\quad-Q^{Z,T}_{\mu'_A\mu'_B}f(E^{Z,T}_{\mu'_A\mu'_B})H(E^{Z,T}_{\mu'_A\mu'_B})\\
\quad\quad\quad\quad-Q^{Z,NT}_{\mu'_A\mu'_B}f(E^{Z,NT}_{\mu'_A\mu'_B})H(E^{Z,NT}_{\mu'_A\mu'_B}),
\end{array}
\end{equation}
where $Q^{Z,NT}_{\mu'_A\mu'_B}$ and $E^{Z,NT}_{\mu'_A\mu'_B}$ is the overall gain and quantum bit error rate of non-triggered events in the $Z$ basis.

After all, the final secret key rate of the modified passive three-intensity decoy-state protocol is given by
\begin{equation}
\label{M-Passive-8}
\begin{array}{lll}
R=\max\{R^{T},R^{Both}\}.
\end{array}
\end{equation}

\section{numerical simulation}
In this section, we numerically compare the performance for the phase-encoded MDI-QKD scheme using the active three-intensity decoy-state protocol, passive two-intensity decoy-state protocol and modified passive three-intensity decoy-state protocol. The numerical parameters used are listed in Table I.
\begin{table}[!b]
\label{Table I}
\caption{List of experimental parameters used in the simulations: $\alpha$ is the loss coefficient of the channel(fiber), $f$ is the error correction inefficiency, $\eta_D$ is the detection efficiency of the relay, $e_D$ is the errors due to channel relative-phase misalignment between Alice and Bob, $p_d$ is the background dark count rate of the detector in the relay.}
\tabcolsep=11pt
\begin{tabular*}{86mm}{lcccc}\hline\hline
$\alpha$(dB/km) & $f$  & $\eta_D$ & $e_d$ & $p_d$\\ \hline
0.20            & 1.16 &  0.145   & 0.015 & $3\times10^{(-6)}$\\ \hline\hline
\end{tabular*}
\end{table}
\begin{figure}[!b]\center
\resizebox{8.5cm}{!}{
\includegraphics{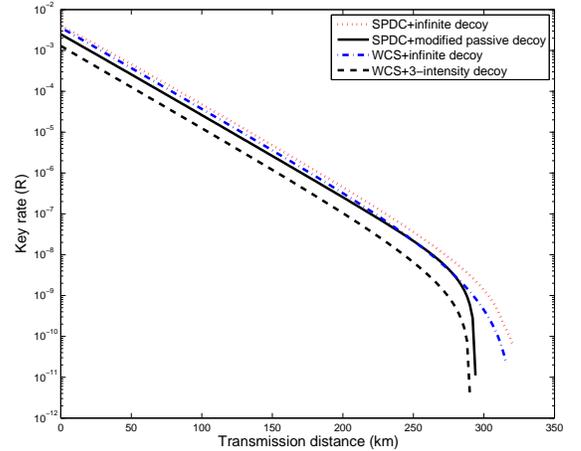}}
\caption{(Color online) Key rate comparison for phase-encoded MDI-QKD using WCS and SPDCS. The setup parameters are listed in Table I. The dotted curve, obtained from Eq.(\ref{KRT-4},\ref{KRT-5}), represents the infinite decoy-state protocol using SPDCS with PND given by Eq.(\ref{HSPS-C}) The dash-dotted curve, obtained from Eq.(\ref{KRT-4}) and Eq.(2-7) in Ref.(\cite{MDI-d1}, represents the infinite decoy-state protocol using WCS). The solid curve, obtained from Eq.(\ref{M-Passive-4},\ref{M-Passive-5},\ref{M-Passive-6},\ref{M-Passive-7},\ref{M-Passive-8}), represents the modified passive three-intensity decoy protocol using SPDCS with PND given by Eq.(\ref{Passive-01},\ref{Passive-02}), where the intensities are chosen by $u_A=u_B=147.577\times10^{-3}, P_{\mu_A}^{cor}=P_{\mu_B}^{cor}=0.12$ and  $u'_A=u'_B=623.927\times10^{-3}, P_{\mu'_A}^{cor}=P_{\mu'_B}^{cor}=0.1$. The dashed curve represents the active three-intensity decoy protocol using WCS, which is obtained from Eq.(\ref{active-1},\ref{active-2},\ref{KRT-4}), where the intensities are chosen by $u_A=u_B=0.2$ and $u'_A=u'_B=0.5$. The related formulas for overall gain and QBER can be found in appendix A and appendix B.}
\end{figure}
It should be noted that all these parameters are the same as Ref.(\cite{MDI1,MDI5,MDI-d1}) for well comparison. The parameters for SPDCS are borrowed from the experiment by Wang et al.\cite{Trigger1,HSPS2}, i.e. $d_A=d_B=5\times10^{(-5)}$, $\eta_A=\eta_B=0.4$ and $P^{cor}_{\mu_A}=P^{cor}_{\mu_B}=0.405844$. For fair comparison, different decoy-state protocols either for WCS and SPDCS apply the following key rate formula
\begin{equation}
\label{KRT-4}
\begin{array}{lll}
R=Q_0+Q^{Z}_{11}(1-H(E^{X}_{11}))\\
\quad\quad-Q^Z_{\mu_A\mu_B}f(E^Z_{\mu_A\mu_B})H(E^Z_{\mu_A\mu_B})
\end{array}
\end{equation}
where $Q_0=P_0(\mu_A)Q^X_{0\mu_B}+P_0(\mu_B)Q^X_{\mu_A0}-P_0(\mu_A)P_0(\mu_B)Y_{00}$. Note that a successful events in the relay need responses from two detectors, thus $Y_{00}={p_d}^2$. $Q^{X}_{11}$ and $E^{X}_{11}$ subject to different constraints for different decoy-state protocols given the measured values of overall gain and QBER in a certain experiment. For MDI-QKD using SPDCS, since no experiment has been reported up to now, we can not obtain the experimental data about the overall gain and QBER. However, we can theoretically estimate them and a stringent evaluation of the overall gain and QBER with full phase randomization is shown in Appendix A. For the third part on the right hand of Eq.(\ref{KRT-4}), the calculation formulas of $Q^Z_{\mu_A\mu_B}$ and $E^Z_{\mu_A\mu_B}$ for SPDCS are given in Appendix B following the way of Ma's method\cite{MDI5}. It should be noted that, for the ideal infinite decoy-state protocol, $Q^Z_{11}$ and $E^{X}_{11}$ are given by\cite{MDI-d1}
\begin{equation}
\label{KRT-5}
\begin{array}{lll}
Q^Z_{11}=P_1(\mu_A)P_1(\mu_B)(1-p_d)^2[\frac{\eta^c_A\eta^c_B}{2}\\
\quad\quad\quad-(2\eta^c_A+3\eta^c_B-3\eta^c_A\eta^c_B)p_d+4(1-\eta^c_A)(1-\eta^c_B){p_d}^2]\\
E^{X}_{11}=e_0-(e_0-e_d)(1-p_d)^2\frac{P_1(\mu_A)P_1(\mu_B)\eta^c_A\eta^c_B}{2Q^X_{11}}
\end{array}
\end{equation}

\begin{figure}[!t]\center
\resizebox{7cm}{!}{
\includegraphics{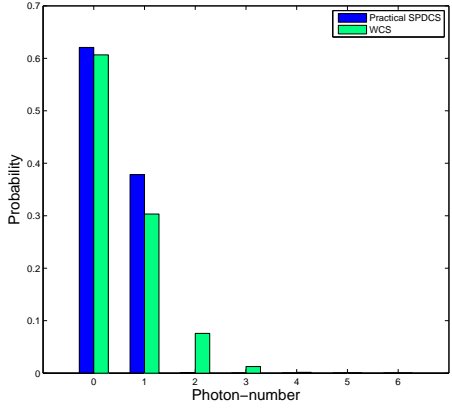}}
\caption{A comparison for the probability of emitting different number of photons between heralded SPDCS and WCS. Here, $P^{cor}_{x}=0.405844$ and $\mu_x=1.425\times10^{-3}$ for the SPDCS while the intensity is $0.5$ for the WCS.}
\end{figure}

In Fig.2, we compare the performance of phase-encoded MDI-QKD using WCS and SPDCS. The dotted curve is with SPDCS and dash-dotted curve is with WCS. Both of them are conducted under the ideal infinite decoy-state protocol and the optimal intensity is selected at each value of transmission distance. The solid curve is conducted under the modified passive three-intensity decoy-state protocol and the dashed curve is obtained under the active three-intensity decoy-state protocol. From Fig.2, it can be seen that the phase-encoded MDI-QKD using SPDCS performs better both in transmission distance and key generation rate than the one using WCS. Our results show that SPDCS shall be superior in the near-future experiment of phase-encoded MDI-QKD. This is because the fraction of single photon for the heralded SPDCS is bigger than that for the WCS. For example, the probability of single photon for the distribution shown in Eq.(\ref{HSPS-C}) is $0.3784$ when $P^{cor}_{x}=0.405844$ and $\mu_x=1.425\times10^{-3}$. But that for the Poisson distribution is 0.3033 when the intensity of WCS is $0.5$. For better interpret this fact, we take a simple comparison in Fig.3. Besides, from our simulation, one can find that the modified passive three-intensity decoy-state protocol and active three-intensity decoy-state protocol are both close to the ideal infinite decoy-state protocol, which implies they are so efficient that can be used in practical experiment.

In Fig.4, we compare the key generation rate of the modified passive three-intensity decoy-state protocol, active three-intensity decoy-state protocol and passive two-intensity decoy-state protocol while practical SPDCS are used in MDI-QKD. From Fig.4, the first two protocols perform better than the third one and are more close to the ideal infinite decoy-state protocol. The result comes from the inefficient estimation of single photon contribution and single photon error in passive two-intensity decoy-state protocol. Most importantly, by introducing one more intensity, our improved passive protocol(the modified passive three-intensity decoy-state protocol) can perform better in transmission distance and key generation rate than the active three-intensity decoy-state protocol. This is resulted from the fact that, combining the events from pulses of another intensity (active decoy sate) and those from non-triggered pulses, our estimations for the single photon yield and error rate are more stringent.

\begin{figure}[!t]\center
\resizebox{8.5cm}{!}{
\includegraphics{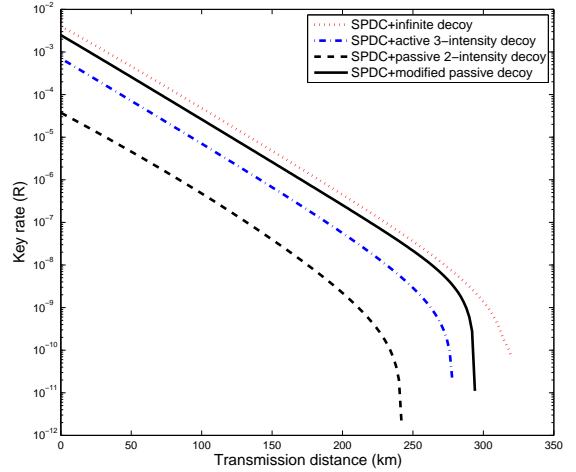}}
\caption{(Color online) Key rate comparison among the modified passive three-intensity decoy-state protocol, active three-intensity decoy-state protocol and passive two-intensity decoy-state protocol for MDI-QKD using practical SPDCS. The setup parameters are listed in Table I. The solid and dotted curve are plotted the same as Fig.2.  The dash-dotted curve, which is obtained from Eq.(\ref{active-1},\ref{active-2},\ref{KRT-4}), represents the active three-intensity decoy protocol using SPDCS with PND given by Eq.(\ref{HSPS-C}), where the intensities are chosen by $u_A=u_B=0.577\times10^{-3}, P_{\mu_A}^{cor}=P_{\mu_B}^{cor}=0.432837$ and  $u'_A=u'_B=1.425\times10^{-3}, P_{\mu'_A}^{cor}=P_{\mu'_B}^{cor}=0.405844$. The dashed curve, plotted referring to Eq.(\ref{Passive-3},\ref{Passive-6},\ref{Passive-7},\ref{Passive-8},\ref{Passive-9}), denotes the passive two-intensity decoy-state protocol using SPDCS with PND given by Eq.(\ref{Passive-01},\ref{Passive-02}), where the intensities are chosen by $u_A=u_B=0.577\times10^{-3}, P_{\mu_A}^{cor}=P_{\mu_B}^{cor}=0.432837$ and  $u'_A=u'_B=790\times10^{-3}, P_{\mu'_A}^{cor}=P_{\mu'_B}^{cor}=0.1$. The related formulas for overall gain and QBER can be found in appendix A and appendix B.}
\end{figure}

\section{statistical fluctuation}
In real-life situations, one cannot generate an infinite number of signals in a reasonable experimental time. Thus the effect of finite data-set size, which will induce statistical fluctuation, shall be included in parameter estimation. In this paper, we apply a rough finite-key analysis based on the standard statistical analysis\cite{Prac-Decoy} and a rigorous analysis might be considered in our future work. With standard error analysis using essentially normal distributions\cite{MDI-d1}, we take a upper and lower bound to characterize the statistical fluctuations of the overall gain and quantum bit error rate, which are given by
\begin{equation}
\label{Fluctuation-1}
\begin{array}{lll}
{\hat{Q}}^{\omega}_{\mu_A\mu_B}(1-\beta_{q})\leq Q^{\omega}_{\mu_A\mu_B}\leq {\hat{Q}}^{\omega}_{\mu_A\mu_B}(1+\beta_{q})\\
{\hat{E}}^{\omega}_{\mu_A\mu_B}{\hat{Q}}^{\omega}_{\mu_A\mu_B}(1-\beta_{eq})\leq Q^{\omega}_{\mu_A\mu_B} E^{\omega}_{\mu_A\mu_B}\\
\quad\quad\quad\quad\quad\quad\quad\quad\quad\quad\leq {\hat{E}}^{\omega}_{\mu_A\mu_B}{\hat{Q}}^{\omega}_{\mu_A\mu_B}(1+\beta_{eq})
\end{array}
\end{equation}
where ${\hat{Q}}^{\omega}_{\mu_A\mu_B}$ and ${\hat{E}}^{\omega}_{\mu_A\mu_B}$ are measurement outcomes in the $\omega$ basis, which can be theoretically estimated in appendix A in our paper. The fluctuation ratio $\beta_{q}$ and $\beta_{eq}$ satisfy $\beta_{q}=n_\alpha/\sqrt{N^{\omega}_{\mu_A\mu_B}{\hat{Q}}^{\omega}_{\mu_A\mu_B}}$ and $\beta_{eq}=n_\alpha/\sqrt{N^{\omega}_{\mu_A\mu_B}{\hat{Q}}^{\omega}_{\mu_A\mu_B}E^{\omega}_{\mu_A\mu_B}}$. Here, $N^{\omega}_{\mu_A\mu_B}$ is the number of pulses, in the $\omega$ basis, sent out by Alice and Bob when they choose intensities $\mu_A$ and $\mu_B$ respectively; $n_\alpha$ is the number of standard deviations one chooses for statistical fluctuation analysis, which is related to the failure probability of the security analysis. Take $n_\alpha=5$ for an example, which represents that the failure probability is $5.73\times10^{-7}$.

If we consider the modified passive three-intensity decoy-state protocol, the bound on the single-photon yield and error rate should be rewritten by
\begin{equation}
\label{Fluctuation-2}
\begin{array}{lll}
Y^Z_{11}\geq \underline{Y^Z_{11}}\\ \quad\quad\equiv\frac{P_1^{T}(\mu_A)P_2^{T}(\mu_B)(\underline{Q_{\mu'_A\mu'_B}^{NT}}-\overline{\tilde{Q}_0^{NT}})}{P_1^{NT}(\mu'_A)P_1^{T}(\mu_A)[P_2^{T}(\mu_B)P_1^{NT}(\mu'_B)-P_2^{NT}(\mu'_B)P_1^{T}(\mu_B)]}\\
\quad\quad\quad-\frac{P_1^{NT}(\mu'_A)P_2^{NT}(\mu'_B)(\overline{Q_{\mu_A\mu_B}^{T}}-\underline{\tilde{Q}_0^{T}})}{P_1^{NT}(\mu'_A)P_1^{T}(\mu_A)[P_2^{T}(\mu_B)P_1^{NT}(\mu'_B)-P_2^{NT}(\mu'_B)P_1^{T}(\mu_B)]},
\end{array}
\end{equation}
\begin{equation}
\label{Fluctuation-3}
\begin{array}{lll}
E^X_{11}\leq \overline{E^X_{11}}\\ \quad\quad\equiv\frac{\overline{E^T_{\mu_A\mu_B}Q^T_{\mu_A\mu_B}}-P_0^T(\mu_A)\underline{E^T_{0\mu_B}Q^T_{0\mu_B}}-P_0^T(\mu_B)\underline{E^T_{\mu_A0}Q^T_{\mu_A0}}}{P_1^T(\mu_A)P_1^T(\mu_B)\underline{Y^Z_{11}}}\\
\quad\quad\quad+\frac{P_0^T(\mu_A)P_0^T(\mu_B)E_{00}Y_{00}}{P_1^T(\mu_A)P_1^T(\mu_B)\underline{Y^Z_{11}}},
\end{array}
\end{equation}
where
\begin{equation}
\label{Fluctuation-4}
\begin{array}{lll}
\overline{Q_{\mu_A\mu_B}^{T}}=Q_{\mu_A\mu_B}^{T}(1+\frac{n_\alpha}{\sqrt{N^{T}_{\mu_A\mu_B}Q^{T}_{\mu_A\mu_B}}}),\\ \underline{Q_{\mu'_A\mu'_B}^{NT}}=Q_{\mu'_A\mu'_B}^{NT}(1-\frac{n_\alpha}{\sqrt{N^{NT}_{\mu'_A\mu'_B}Q^{NT}_{\mu'_A\mu'_B}}}),\\
\overline{E^T_{\mu_A\mu_B}Q^T_{\mu_A\mu_B}}=E^T_{\mu_A\mu_B}Q^T_{\mu_A\mu_B}(1+\frac{n_\alpha}{\sqrt{N^{T}_{\mu_A\mu_B}Q^{T}_{\mu_A\mu_B}E^{T}_{\mu_A\mu_B}}}),\\ \quad\underline{E^T_{0\mu_B}Q^T_{0\mu_B}}=E^T_{0\mu_B}Q^T_{0\mu_B}(1-\frac{n_\alpha}{\sqrt{N^{T}_{0\mu_B}Q^{T}_{0\mu_B}E^{T}_{0\mu_B}}}),\\
\quad\underline{E^T_{\mu_A0}Q^T_{\mu_A0}}=E^T_{\mu_A0}Q^T_{\mu_A0}(1-\frac{n_\alpha}{\sqrt{N^{T}_{\mu_A0}Q^{T}_{\mu_A0}E^{T}_{\mu_A0}}}).
\end{array}
\end{equation}
\begin{figure}[!t]\center
\resizebox{8.5cm}{!}{
\includegraphics{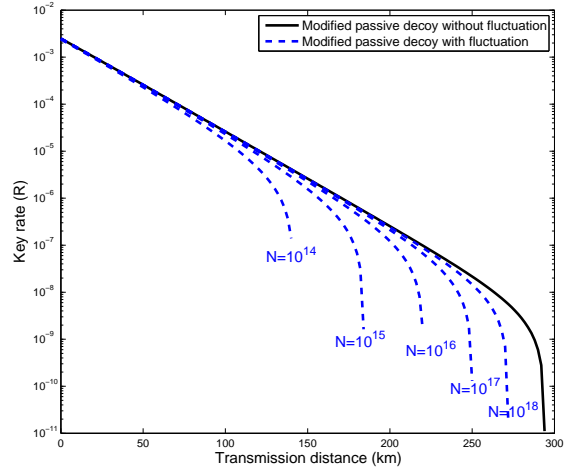}}
\caption{(Color online) Key rate of MDI-QKD with statistical fluctuation under the modified passive three-intensity decoy-state protocol. The setup parameters are listed in Table I. The solid curve is plotted the same as Fig.2. The dashed curve is obtained from Eq.(\ref{Fluctuation-2},\ref{Fluctuation-3},\ref{M-Passive-6},\ref{M-Passive-7},\ref{M-Passive-8}) for our modified passive three-intensity decoy-state protocol with different length of data $N$ . The related formulas for overall gain and QBER can be found in appendix A and appendix B. In the simulations, we assume that five standard deviations ($n_\alpha=5$) are used.}
\end{figure}
In the above equations, $\underline{\tilde{Q}_0^{T}}={P_{0}^{T}}(\mu_B)\underline{Q_{\mu_A0}^{T}}+{P_{0}^{T}}(\mu_A)\underline{Q_{0\mu_B}^{T}}-Q_{00}^{T}$, $\overline{\tilde{Q}_0^{NT}}={P_{0}^{NT}}(\mu_B')\overline{Q_{\mu_A'0}^{NT}}+{P_{0}^{NT}}(\mu_A')\overline{Q_{0\mu_B'}^{NT}}-Q_{00}^{NT}$; $\underline{Q_{\mu_A0}^{T}}=Q_{\mu_A0}^{T}(1-\frac{n_\alpha}{\sqrt{N^{T}_{\mu_A0}Q^{T}_{\mu_A0}}})$,
$\underline{Q_{0\mu_B}^{T}}=Q_{0\mu_B}^{T}(1-\frac{n_\alpha}{\sqrt{N^{T}_{0\mu_B}Q^{T}_{0\mu_B}}})$, $\overline{Q_{\mu'_A0}^{NT}}=Q_{\mu'_A0}^{NT}(1+\frac{n_\alpha}{\sqrt{N^{NT}_{\mu'_A0}Q^{NT}_{\mu'_A0}}})$, $\overline{Q_{0\mu'_B}^{NT}}=Q_{0\mu'_B}^{NT}(1+\frac{n_\alpha}{\sqrt{N^{NT}_{0\mu'_B}Q^{NT}_{0\mu'_B}}})$;
$N^{T}_{\mu_A\mu_B}$, $N^{T}_{\mu_A0}$ and $N^{T}_{0\mu_B}$ denote the number of the triggered pulses when Alice and Bob choose intensities of $\{\mu_A\mu_B\}$, $\{\mu_A0\}$, and $\{0\mu_B\}$ respectively.  $N^{NT}_{\mu'_A\mu'_B}$, $N^{NT}_{\mu'_A0}$ and $N^{NT}_{0\mu'_B}$ denote the number of the non-triggered pulses when Alice and Bob choose intensities of $\{\mu'_A\mu'_B\}$, $\{\mu'_A0\}$, and $\{0\mu'_B\}$ respectively.

Substituting Eq.(\ref{Fluctuation-2}) and Eq.(\ref{Fluctuation-3}) into Eq.(\ref{M-Passive-6}) and Eq.(\ref{M-Passive-7}), we can
estimate the secret key rate with data under statistical fluctuation, which is shown in Fig.5. In the simulations, we choose standard deviations with $n_\alpha=5$ and the length of pulses is assumed to be the same for each pair of intensities of Alice and Bob. Particularly, if we define $N_{\mu_A\mu_B}$, $N_{\mu'_A\mu'_B}$, $N_{\mu_A0}$, $N_{\mu'_A0}$, $N_{0\mu_B}$ and $N_{0\mu'_B}$ as the number of pulses when Alice and Bob choose different intensities and assume $N_{\mu_A\mu_B}=N_{\mu'_A\mu'_B}=N_{\mu_A0}=N_{\mu'_A0}=N_{0\mu_B}=N_{0\mu'_B}\triangleq N$, then the unknown variables in Eq.(\ref{Fluctuation-2}) and Eq.(\ref{Fluctuation-3}) can be obtained by $N^{T}_{\mu_A\mu_B}=\eta_A\eta_BN$, $N^{NT}_{\mu'_A\mu'_B}=(1-\eta_A)(1-\eta_B)N$, $N^{T}_{\mu_A0}=\eta_Ad_BN$, $N^{NT}_{\mu'_A0}=(1-\eta_A)(1-d_B)N$, $N^{T}_{0\mu_B}=d_A\eta_BN$ and $N^{NT}_{0\mu'_B}=(1-d_A)(1-\eta_B)N$. From Fig.5, one can clearly find that the finite length of the raw key will obviously compromise the secret key rate of our protocol and large number of signals is required for preferable secret key being shared by two parties at a long distance. A rigorous finite-key analysis for our protocol requires further study.

\section{conclusion}
In conclusion, we put forward two passive decoy-state protocol for MDI-QKD using practical SPDCS and analyze the performance of it under different decoy-state protocols. From numerical simulation, we remark that the MDI-QKD using practical SPDCS performs better than the one using WCS. Furthermore, we conclude that the modified passive three-intensity decoy-state protocol we proposed is close to the ideal infinite decoy-state protocol and shall be as a choice for the practical experiment of MDI-QKD using SPDCS.

For a complete security analysis, issues like the finite key effect, intensity fluctuation and modulator imperfection, must also be considered. Curty et al.\cite{Finite-MDI} recently gives a complete and rigorous finite-key analysis on MDI-QKD using WCS. It will be attractive to analyze these issues for the MDI-QKD using practical SPDCS. In this paper, we assume the devices and transmission distance are the same for the two senders. It is interesting to have a research on the case when two source settings are different and recent study by Xu et al.\cite{Prac-MDI} has an important progress about this issue.
\section*{ACKNOWLEDGMENTS}
The authors would like to thank Xiongfeng Ma, Jingzheng Huang and Shuang Wang for their helpful discussions and, in particular, the referees for providing enlightening ideas and giving important suggestions. The authors gratefully acknowledge the financial support from the National Basic Research Program of China (Grants No. 2011CBA00200 and No. 2011CB921200), the National Natural Science Foundation of China (Grant Nos. 61101137, 61201239, 61205118).

\section*{APPENDIX A: OVERALL QUANTUM BIT GAIN AND QUANTUM BIT ERROR RATE WITH FULL PHASE RANDOMIZATION}
In practical fiber-based QKD experiment, the overall gain and error rate can be directly measured. However, in this appendix, we shall give their theoretical evaluation for simulation purpose. In what follows, we will evaluate the overall gain and QBER for the PND under the condition that the detector is triggered or not triggered. We shall consider the case when the signal states emitted from the SPDCS are phase-randomized.

First of all, note the fact that the degree of influence of fibre loss on the transmitted mixed state is different for different type of PND, we shall characterize how fibre loss affects the transmitted  mixed state under a practical PND. Interestingly, the basic setup of one type of PND passing through a beam splitter can be introduced to characterize this influence.
\begin{figure}[!h]\center
\resizebox{6cm}{!}{
\includegraphics{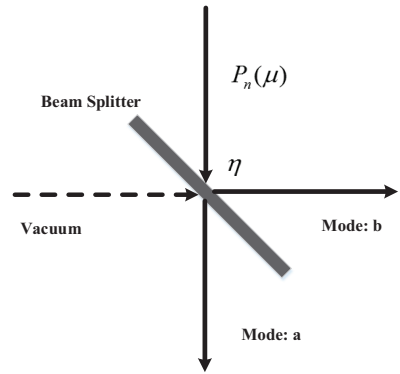}}
\caption{Diagram of basic setup for a photon-number distribution passing through a beams splitter with transmittancy of $\eta$}
\end{figure}
In this scenario, as shown in Fig.6 when one state with number distribution of $P_n(\mu)$ passing through a beam splitter with transmittancy $\eta$, the photon-number distribution ${P}_{n,m}(\mu)$ of events having n photons in output mode a and m photons in output mode b is a Bernoulli transform of the distribution $P_n(\mu)$ about $\eta$, that is,
\begin{equation}
\begin{array}{lll}
{P}_{n,m}(\mu)=C_{n+m}^{m}{{P}_{n+m}}(\mu ){{\eta }^{n}}{(1-\eta )}^{m}.
\end{array}\\\tag{A1}
\end{equation}
In particular, when one ignores the output of mode b, the conditional photon-number distribution $P_n(\mu,\eta)$ of output mode a is the total probability over all region of m, which can be expressed as
\begin{equation}
\begin{array}{lll}
{{P}_{n}}(\mu,\eta)=\sum\limits_{m=0}^{\infty }{{{P}_{n,m}}(\mu)}\\\quad\quad\quad\quad=\sum\limits_{m=0}^{\infty}{C_{n+m}^{m}{{P}_{n+m}}(\mu){{\eta }^{n}}{(1-\eta)}^{m}}.
\end{array}\\\tag{A2}
\end{equation}
With mathematical Taylor expansion, it is easy to prove that the equation ${{\alpha }^{-n-1}}=\sum\limits_{m=0}^{\infty }{{{(1-\alpha )}^{m}}C_{n+m}^{m}}$ always holds true for any real number $0<\alpha<1$. Thus, for a thermal photon-number distribution with $P_n(\mu)=\frac{\mu^n}{(1+\mu)^{n+1}}$, the number distribution after the beams splitter is
\begin{equation}
\begin{array}{lll}
{{P}_{n}}(\mu ,\eta)=\frac{(\mu \eta )^{n}}{(1+\mu \eta )^{n+1}}.
\end{array}\\\tag{A3}
\end{equation}
Under this method, we can obtain the formalization of the triggered mixed state, given by Eq.(\ref{HSPS-A})and Eq.(\ref{HSPS-B}), after passing through the quantum channel with transmittancy of $\eta^c_A$($\eta^c_B$)for Alice(Bob), that is
\begin{equation}
\label{HSPS-1}
\begin{array}{lll}
{{\left|\psi({{\mu}_{x}},{{\theta}_{x}},{\eta^c_x})\ \right\rangle}_{x}}\\
=\sqrt{\frac{P_{\mu_x}^{cor}}{P_{\mu_x}^{post}}(\frac{1+d_x}{1+\mu_x\eta^c_x}-\frac{1}{1+\mu_x(\eta^c_x+\eta_x-\eta_x\eta^c_x)})+(1-P_{\mu_x}^{cor})}\left|0\right\rangle\\
\quad+\sum\limits_{n=1}^{\infty}{\sqrt{\frac{P_{\mu_x}^{cor}}{P_{\mu_x}^{post}}F_n(\mu_x,\eta^c_x)}{{e}^{in\theta_x}}\left|n\right\rangle}
\\\\with \\ F_n(\mu_x,\eta^c_x)=\frac{(1+{{d}_{x}}){{({{\mu}_{x}}{\eta^c_x})}^{n}}}{{{(1+{{\mu}_{x}}{\eta^c_x})}^{n+1}}}-\frac{{{\left[{{\mu}_{x}}{\eta^c_x}(1-{{\eta}_{x}}) \right]}^{n}}}{{{\left[1+{{\mu}_{x}}({\eta^c_x}+{{\eta}_{x}}-{{\eta }_{x}}{\eta^c_x})\right]}^{n+1}}},
\end{array}\\\tag{A4}
\end{equation}
where $x=A$ or $x=B$.

Alice and Bob prepare the following heralded mixed states with intensities $\mu_A$ and $\mu_B$ respectively before phase randomization
\begin{equation}
\label{HSPS-G}
\begin{array}{lll}
{\left|\psi(\mu_A,\theta_A)\ \right\rangle}_A {\left|\psi(\mu_B,\theta_B)\ \right\rangle }_B.
\end{array}\\\tag{A5}
\end{equation}
Emitted from the sources, the joint photon state are split by 50:50 beam splitters, labeled by $r$ and $s$, and transmitted through the lossy channel. Then, the state arrived at the relay can be described by
\begin{equation}
\label{HSPS-H}
\begin{array}{lll}
{{\left|\psi(\mu_A,\theta_A,\frac{\eta^c_A}{2})\ \right\rangle}_{A_r}} {{\left|\psi(\mu_A,\theta_A+\phi_A,\frac{\eta^c_A}{2})\ \right\rangle}_{A_s}}
\\\otimes{{\left|\psi(\mu_B,\theta_B,\frac{\eta^c_B}{2})\ \right\rangle}_{B_r}} {{\left|\psi(\mu_B,\theta_B+\phi_B,\frac{\eta^c_B}{2})\ \right\rangle}_{B_s}}.
\end{array}\\\tag{A6}
\end{equation}
where $\phi_A$ and $\phi_B$ are the encoded information of phase that Alice and Bob want to apply. After going through the beam splitters in the relay, the above state will be transformed into the following partial Bell-state measurement detection modes $r_0$, $r_1$, $r_2$ and $r_3$.
\begin{equation}
\label{HSPS-I}
\begin{array}{lll}
{{\left|\frac{1}{\sqrt{2}}\psi(\mu_A,\theta_A,\frac{\eta^c_A}{2})+\frac{1}{\sqrt{2}}\psi(\mu_B,\theta_B,\frac{\eta^c_B}{2})\ \right\rangle}_{r_0}} \\\otimes
{{\left|\frac{1}{\sqrt{2}}\psi(\mu_A,\theta_A,\frac{\eta^c_A}{2})-\frac{1}{\sqrt{2}}\psi(\mu_B,\theta_B,\frac{\eta^c_B}{2})\ \right\rangle}_{r_1}}
\\\otimes
{{\left|\frac{1}{\sqrt{2}}\psi(\mu_A,\theta_A+\phi_A,\frac{\eta^c_A}{2})+\frac{1}{\sqrt{2}}\psi(\mu_B,\theta_B+\phi_B,\frac{\eta^c_B}{2})\ \right\rangle}_{s_0}}
\\\otimes
{{\left|\frac{1}{\sqrt{2}}\psi(\mu_A,\theta_A+\phi_A,\frac{\eta^c_A}{2})-\frac{1}{\sqrt{2}}\psi(\mu_B,\theta_B+\phi_B,\frac{\eta^c_B}{2})\ \right\rangle}_{s_1}}.
\end{array}\\\tag{A7}
\end{equation}
In what follows, we shall calculate the detection probabilities for these four detectors. For simplification, we denote
\begin{equation}
\label{HSPS-J}
\begin{array}{lll}
a_0=1-P_{\mu_A}^{cor}\\
\quad\quad+\frac{P_{\mu_A}^{cor}}{P_{\mu_A}^{post}}(\frac{1+d_A}{1+0.5\mu_A\eta^c_A}-\frac{1}{1+\mu_A(0.5\eta^c_A+\eta_A-0.5\eta_A\eta^c_A)})\\
b_0=1-P_{\mu_B}^{cor}\\
\quad\quad+\frac{P_{\mu_B}^{cor}}{P_{\mu_B}^{post}}(\frac{1+d_B}{1+0.5\mu_B\eta^c_B}-\frac{1}{1+\mu_B(0.5\eta^c_B+\eta_B-0.5\eta_B\eta^c_B)})\\
a_n=\frac{P_{{{\mu}_{A}}}^{cor}}{P_{\mu_A}^{post}}F_n(\mu_A,\frac{\eta^c_A}{2})\\
=\frac{P_{{{\mu}_{A}}}^{cor}}{P_{\mu_A}^{post}}\{{\frac{(1+d_A){(\frac{\mu_A\eta^c_A}{2})}^{n}}{{(1+\frac{\mu_A\eta^c_A}{2})}^{n+1}}-\frac{{{\left[\frac{\mu_A\eta^c_A}{2}(1-\eta_A)\right]}^n}}{{{\left[1+\mu_A(\frac{\eta^c_A}{2}+\eta_A-\frac{\eta_A\eta^c_A}{2})\right]}^{n+1}}}}\}\\
b_n=\frac{P_{{{\mu}_{B}}}^{cor}}{P_{\mu_B}^{post}}F_n(\mu_B,\frac{\eta^c_B}{2})\\
=\frac{P_{{{\mu}_{B}}}^{cor}}{P_{\mu_B}^{post}}\{{\frac{(1+d_B){(\frac{\mu_B\eta^c_B}{2})}^{n}}{{(1+\frac{\mu_B\eta^c_B}{2})}^{n+1}}-\frac{{{\left[\frac{\mu_B\eta^c_B}{2}(1-\eta_B)\right]}^n}}{{{\left[1+\mu_B(\frac{\eta^c_B}{2}+\eta_B-\frac{\eta_B\eta^c_B}{2})\right]}^{n+1}}}}\}
\end{array}\\\tag{A8}
\end{equation}
Let $\sqrt{T_0}=\sqrt{\frac{a_0}{2}}+\sqrt{\frac{b_0}{2}}$ and $\sqrt{K_0}=\sqrt{\frac{a_0}{2}}-\sqrt{\frac{b_0}{2}}$. Then, Eq.(\ref{HSPS-I}) can be expressed as
\begin{equation}
\label{HSPS-K}
\begin{array}{lll}
[\sqrt{T_0}\left|0\right\rangle+\sum\limits_{n=1}^{\infty}{(\sqrt{\frac{a_n}{2}}{e}^{in\theta_A}+\sqrt{\frac{b_n}{2}}{e}^{in\theta_B})\left|n\right\rangle}]_{r_0} \\\otimes
[\sqrt{K_0}\left|0\right\rangle+\sum\limits_{n=1}^{\infty}{(\sqrt{\frac{a_n}{2}}{e}^{in\theta_A}-\sqrt{\frac{b_n}{2}}{e}^{in\theta_B})\left|n\right\rangle}]_{r_1} \\\otimes
[\sqrt{T_0}\left|0\right\rangle+\sum\limits_{n=1}^{\infty}{(\sqrt{\frac{a_n}{2}}{e}^{in(\theta_A+\phi_A)}+\sqrt{\frac{b_n}{2}}{e}^{in(\theta_B+\phi_B)})\left|n\right\rangle}]_{r_2} \\\otimes
[\sqrt{K_0}\left|0\right\rangle+\sum\limits_{n=1}^{\infty}{(\sqrt{\frac{a_n}{2}}{e}^{in(\theta_A+\phi_A)}-\sqrt{\frac{b_n}{2}}{e}^{in(\theta_B+\phi_B)})\left|n\right\rangle}]_{r_3} \end{array}\\\tag{A9}
\end{equation}
Suppose the properties of the four detectors are the same and let $\eta_D$ and $p_d$ denote their detection efficiency and dark count rate respectively. Therefore, for an incoming $n$-photon states, the probability of at least one photon being detected can be described by
\begin{equation}
\label{HSPS-L}
\begin{array}{lll}
D_n=p_d+(1-p_d)[1-{(1-\eta_D)}^n]\\
\quad\quad=1-(1-p_d)(1-\eta_D)^n.
\end{array}\\\tag{A10}
\end{equation}
Thus the detection probabilities for the four detectors are given by
\begin{equation}
\label{HSPS-M}
\begin{array}{lll}
D_{r_0}={\mid\sqrt{T_0}\mid}^2 p_d\\
\quad\quad\quad+\sum\limits_{n=1}^{\infty}{\{[\frac{a_n}{2}+\frac{b_n}{2}+\sqrt{a_nb_n}\cos{(n\Delta_{\theta})}]D_n\}},\\
D_{r_1}={\mid\sqrt{K_0}\mid}^2 p_d\\
\quad\quad\quad+\sum\limits_{n=1}^{\infty}{\{[\frac{a_n}{2}+\frac{b_n}{2}-\sqrt{a_nb_n}\cos{(n\Delta_{\theta})}]D_n\}},\\
D_{s_0}={\mid\sqrt{T_0}\mid}^2 p_d\\
\quad\quad\quad+\sum\limits_{n=1}^{\infty}{\{[\frac{a_n}{2}+\frac{b_n}{2}+\sqrt{a_nb_n}\cos{n(\Delta_{\theta}+\Delta_{\phi})}]D_n]\}},\\
D_{s_1}={\mid\sqrt{K_0}\mid}^2 p_d\\
\quad\quad\quad+\sum\limits_{n=1}^{\infty}{\{[\frac{a_n}{2}+\frac{b_n}{2}-\sqrt{a_nb_n}\cos{n(\Delta_{\theta}+\Delta_{\phi})}]D_n]\}}.
\end{array}\\\tag{A11}
\end{equation}
where $\Delta_{\theta}=\theta_A-\theta_B$ and $\Delta_{\phi}=\phi_A-\phi_B$. \\

In actual experiments of Hong-Ou-Mandel (HOM) effect \cite{HOM} happened in the relay, it is assumed that the single photons entered the 50:50 beam splitter are identical. This can make sure there are no unwanted errors when both single photons are prepared under the same basis. Thus, in this paper, we assume Alice's devices and  Bob's have identical performance, which guarantees indistinguishable photons will be generated from two independent practical SPDC sources and stable HOM interference will be observed in the relay. That is to say, we shall assume $\mu_A=\mu_B$, $P_{\mu_A}^{cor}=P_{\mu_B}^{cor}$, $P_{\mu_A}^{post}=P_{\mu_B}^{post}$, $d_A=d_B$ and $\eta_A=\eta_B$. For better simulation performance, although always not the case in practical MDI-QKD experiments, we also assume the channel transmittances between sources and detectors are identical, i.e., $\eta^c_A=\eta^c_B$. Here, $\eta^c_A=10^{-\alpha l_{AC}/{10}}$ and $\eta^c_B=10^{-\alpha l_{BC}/{10}}$. In the above Equation, $\alpha$ denotes the loss coefficient, $l_{AC}$ ($l_{BC}$) denotes the length of the fiber between Alice(Bob) and Charlie. Under the above condition, it can be obtained that $a_0=b_0$ and $a_n=b_n$ for $ n\geq1$. Therefore, Eq.(\ref{HSPS-M}) can be quantified by
\begin{equation}
\label{HSPS-N}
\begin{array}{lll}
D_{r_0}=2a_0p_d+\frac{P_{\mu_A}^{cor}}{P_{\mu_A}^{post}}\{\frac{x(1+d_A)}{1+x}-\frac{v(1+d_A)(1-p_d)}{(1+x)(1+x\eta_D)}\\
\quad\quad\quad-\frac{u}{z(z-u)}+\frac{w(1-p_d)}{z(z-w)}\\
\quad\quad\quad+\frac{x(1+d_A)}{2(1+x)}[\frac{1}{(1+x)e^{i\Delta_{\theta}}-x}+\frac{e^{i\Delta_{\theta}}}{1+x-xe^{i\Delta_{\theta}}}]\\
\quad\quad\quad-\frac{v(1+d_A)(1-p_d)}{2(1+x)}[\frac{1}{(1+x)e^{i\Delta_{\theta}}-v}+\frac{e^{i\Delta_{\theta}}}{1+x-ve^{i\Delta_{\theta}}}]\\
\quad\quad\quad-\frac{u}{2z}[\frac{1}{ze^{i\Delta_{\theta}}-u}+\frac{e^{i\Delta_{\theta}}}{z-ue^{i\Delta_{\theta}}}]\\
\quad\quad\quad+\frac{w(1-p_d)}{2z}[\frac{1}{ze^{i\Delta_{\theta}}-w}+\frac{e^{i\Delta_{\theta}}}{z-we^{i\Delta_{\theta}}}]\},
\end{array}\\\tag{A12}
\end{equation}
\begin{equation}
\label{HSPS-O}
\begin{array}{lll}
D_{r_1}=\frac{P_{\mu_A}^{cor}}{P_{\mu_A}^{post}}\{\frac{x(1+d_A)}{1+x}-\frac{v(1+d_A)(1-p_d)}{(1+x)(1+x\eta_D)}\\
\quad\quad\quad-\frac{u}{z(z-u)}+\frac{w(1-p_d)}{z(z-w)}\\
\quad\quad\quad-\frac{x(1+d_A)}{2(1+x)}[\frac{1}{(1+x)e^{i\Delta_{\theta}}-x}+\frac{e^{i\Delta_{\theta}}}{1+x-xe^{i\Delta_{\theta}}}]\\
\quad\quad\quad+\frac{v(1+d_A)(1-p_d)}{2(1+x)}[\frac{1}{(1+x)e^{i\Delta_{\theta}}-v}+\frac{e^{i\Delta_{\theta}}}{1+x-ve^{i\Delta_{\theta}}}]\\
\quad\quad\quad+\frac{u}{2z}[\frac{1}{ze^{i\Delta_{\theta}}-u}+\frac{e^{i\Delta_{\theta}}}{z-ue^{i\Delta_{\theta}}}]\\
\quad\quad\quad-\frac{w(1-p_d)}{2z}[\frac{1}{ze^{i\Delta_{\theta}}-w}+\frac{e^{i\Delta_{\theta}}}{z-we^{i\Delta_{\theta}}}]\},
\end{array}\\\tag{A13}
\end{equation}
\begin{equation}
\label{HSPS-P}
\begin{array}{lll}
D_{s_0}=2a_0p_d+\frac{P_{\mu_A}^{cor}}{P_{\mu_A}^{post}}\{\frac{x(1+d_A)}{1+x}-\frac{v(1+d_A)(1-p_d)}{(1+x)(1+x\eta_D)}\\
\quad\quad\quad-\frac{u}{z(z-u)}+\frac{w(1-p_d)}{z(z-w)}\\
\quad\quad\quad+\frac{x(1+d_A)}{2(1+x)}[\frac{1}{(1+x)e^{i(\Delta_{\theta}+\Delta_{\phi})}-x}+\frac{e^{i(\Delta_{\theta}+\Delta_{\phi})}}{1+x-xe^{i(\Delta_{\theta}+\Delta_{\phi})}}]\\
\quad\quad\quad-\frac{v(1+d_A)(1-p_d)}{2(1+x)}[\frac{1}{(1+x)e^{i(\Delta_{\theta}+\Delta_{\phi})}-v}+\frac{e^{i(\Delta_{\theta}+\Delta_{\phi})}}{1+x-ve^{i(\Delta_{\theta}+\Delta_{\phi})}}]\\
\quad\quad\quad-\frac{u}{2z}[\frac{1}{ze^{i(\Delta_{\theta}+\Delta_{\phi})}-u}+\frac{e^{i(\Delta_{\theta}+\Delta_{\phi})}}{z-ue^{i(\Delta_{\theta}+\Delta_{\phi})}}]\\
\quad\quad\quad+\frac{w(1-p_d)}{2z}[\frac{1}{ze^{i(\Delta_{\theta}+\Delta_{\phi})}-w}+\frac{e^{i(\Delta_{\theta}+\Delta_{\phi})}}{z-we^{i(\Delta_{\theta}+\Delta_{\phi})}}]\},
\end{array}\\\tag{A14}
\end{equation}
\begin{equation}
\label{HSPS-Q}
\begin{array}{lll}
D_{s_1}=\frac{P_{\mu_A}^{cor}}{P_{\mu_A}^{post}}\{\frac{x(1+d_A)}{1+x}-\frac{v(1+d_A)(1-p_d)}{(1+x)(1+x\eta_D)}\\
\quad\quad\quad-\frac{u}{z(z-u)}+\frac{w(1-p_d)}{z(z-w)}\\
\quad\quad\quad-\frac{x(1+d_A)}{2(1+x)}[\frac{1}{(1+x)e^{i(\Delta_{\theta}+\Delta_{\phi})}-x}+\frac{e^{i(\Delta_{\theta}+\Delta_{\phi})}}{1+x-xe^{i(\Delta_{\theta}+\Delta_{\phi})}}]\\
\quad\quad\quad+\frac{v(1+d_A)(1-p_d)}{2(1+x)}[\frac{1}{(1+x)e^{i(\Delta_{\theta}+\Delta_{\phi})}-v}+\frac{e^{i(\Delta_{\theta}+\Delta_{\phi})}}{1+x-ve^{i(\Delta_{\theta}+\Delta_{\phi})}}]\\
\quad\quad\quad+\frac{u}{2z}[\frac{1}{ze^{i(\Delta_{\theta}+\Delta_{\phi})}-u}+\frac{e^{i(\Delta_{\theta}+\Delta_{\phi})}}{z-ue^{i(\Delta_{\theta}+\Delta_{\phi})}}]\\
\quad\quad\quad-\frac{w(1-p_d)}{2z}[\frac{1}{ze^{i(\Delta_{\theta}+\Delta_{\phi})}-w}+\frac{e^{i(\Delta_{\theta}+\Delta_{\phi})}}{z-we^{i(\Delta_{\theta}+\Delta_{\phi})}}]\},
\end{array}\\\tag{A15}
\end{equation}
where $x=\frac{\mu_{A}\eta_{A}^{c}}{2}$, $z=1+x+\mu_A\eta_A-x\eta_A$, $u=x(1-\eta_A)$, $v=x(1-\eta_D)$ and $w=u(1-\eta_D)=v(1-\eta_A)$. Note the fact that $\Delta_{\phi}=0,\pi$ when Alice and Bob choose the same basis, we take the integral of $\Delta_{\theta}$ for Eq.(\ref{HSPS-N}), Eq.(\ref{HSPS-O}), Eq.(\ref{HSPS-P}) and Eq.(\ref{HSPS-Q}). Then, we obtain
\begin{equation}
\label{HSPS-R}
\begin{array}{lll}
D_{r_0}=D_{s_0}=2a_0p_d+\frac{P_{\mu_A}^{cor}}{P_{\mu_A}^{post}}\{\frac{x(1+d_A)}{1+x}-\frac{v(1+d_A)(1-p_d)}{(1+x)(1+x\eta_D)}\\
\quad\quad\quad\quad\quad\quad-\frac{u}{z(z-u)}+\frac{w(1-p_d)}{z(z-w)}-\frac{(1+d_A)p_d}{2(1+x)}+\frac{p_d}{2z}\},\\
D_{r_1}=D_{s_1}=\frac{P_{\mu_A}^{cor}}{P_{\mu_A}^{post}}\{\frac{x(1+d_A)}{1+x}-\frac{v(1+d_A)(1-p_d)}{(1+x)(1+x\eta_D)}\\
\quad\quad\quad\quad\quad\quad-\frac{u}{z(z-u)}+\frac{w(1-p_d)}{z(z-w)}+\frac{(1+d_A)p_d}{2(1+x)}-\frac{p_d}{2z}\}.
\end{array}\\\tag{A16}
\end{equation}

As is proved by Ma et al.\cite{MDI-d1}, the gain $Q_{\mu_A\mu_B}$ is given by
\begin{equation}
\label{HSPS-S}
\begin{array}{lll}
Q_{\mu_A\mu_B}=[D_{r_0}(1-D_{r_1})+(1-D_{r_0})D_{r_1}][D_{s_0}(1-D_{s_1})\\
\quad\quad\quad\quad+(1-D_{s_0})D_{s_1}].
\end{array}\\\tag{A17}
\end{equation}
We substitute Eq.(\ref{HSPS-R}) into Eq.(\ref{HSPS-S}) and obtain
\begin{equation}
\label{HSPS-T}
\begin{array}{lll}
Q_{\mu_A\mu_B}=[C_0+2PC_1-2C_0P(C_1-C_2)\\
\quad\quad\quad\quad\quad-2P^2({C_1}^2-{C_2}^2)]^2,
\end{array}\\\tag{A18}
\end{equation}
with
\begin{equation}
\label{HSPS-U}
\begin{array}{lll}
C_0=2a_0p_d,\\
P=\frac{P_{\mu_A}^{cor}}{P_{\mu_A}^{post}},\\
C_1=\frac{x(1+d_A)}{1+x}-\frac{v(1+d_A)(1-p_d)}{(1+x)(1+x\eta_D)}-\frac{u}{z(z-u)}+\frac{w(1-p_d)}{z(z-w)},\\
C_2=\frac{p_d}{2z}-\frac{(1+d_A)p_d}{2(1+x)}.
\end{array}\\\tag{A19}
\end{equation}

For the case when $\mu_A=0$ and $\mu_B\neq0$, the calculation formula for $Q_{0\mu_B}$ is an exception of the above Eq.(\ref{HSPS-U}). In that case, Eq.(\ref{HSPS-M}) changes to
\begin{equation}
\label{QBER-1}
\begin{array}{lll}
D_{r_0}=D_{s_0}=\frac{1}{2}\{{\mid(\sqrt{a_0}+\sqrt{b_0})\mid}^2 p_d+\sum\limits_{n=1}^{\infty}{\{b_nD_n\}}\},\\
D_{r_1}=D_{s_1}=\frac{1}{2}\{{\mid(\sqrt{a_0}-\sqrt{b_0})\mid}^2 p_d+\sum\limits_{n=1}^{\infty}{\{b_nD_n\}}\}.
\end{array}\\\tag{A20}
\end{equation}
Here, $a_0=1$ and $b_0=\frac{P_{\mu_B}^{cor}}{P_{\mu_B}^{post}}(\frac{1+d_B}{1+0.5\mu_B\eta^c_B}-\frac{1}{1+\mu_B(0.5\eta^c_B+\eta_B-0.5\eta_B\eta^c_B)})+(1-P_{\mu_B}^{cor})$. With a simple mathematical computation, we can obtain
\begin{equation}
\label{QBER-2}
\begin{array}{lll}
D_{r_0}=D_{s_0}=\frac{1}{2}{\mid(1+\sqrt{b_0})\mid}^2p_d+\frac{P_{\mu_B}^{cor}}{2P_{\mu_B}^{post}}\{\frac{x(1+d_B)}{1+x}\\
\quad\quad\quad\quad\quad\quad-\frac{v(1+d_B)(1-p_d)}{(1+x)(1+x\eta_D)}-\frac{u}{z(z-u)}+\frac{w(1-p_d)}{z(z-w)}\},\\
D_{r_1}=D_{s_1}=\frac{1}{2}{\mid(1-\sqrt{b_0})\mid}^2p_d+\frac{P_{\mu_B}^{cor}}{2P_{\mu_B}^{post}}\{\frac{x(1+d_B)}{1+x}\\
\quad\quad\quad\quad\quad\quad-\frac{v(1+d_B)(1-p_d)}{(1+x)(1+x\eta_D)}-\frac{u}{z(z-u)}+\frac{w(1-p_d)}{z(z-w)}\}.
\end{array}\\\tag{A21}
\end{equation}
Therefore, we can obtain $Q_{0\mu_B}$ according to Eq.(\ref{HSPS-S}). $Q_{\mu_A0}$ can be calculated under the same way.

The intrinsic error rate, resulted form the background noise and multi-photon states, is given by \cite{MDI-d1}
\begin{equation}
\label{QBER-3}
\begin{array}{lll}
E'_{\mu_A\mu_B}Q_{\mu_A\mu_B}=2D_{r_0}(1-D_{r_1})(1-D_{s_1})D_{s_0}.
\end{array}\\\tag{A22}
\end{equation}
Considering the relative-phase distortion errors, we can obtain the averaged quantum bit error rate
\begin{equation}
\label{QBER-10}
\begin{array}{lll}
E_{\mu_A\mu_B}=E'_{\mu_A\mu_B}+e_d(1-\frac{E_{\mu_A\mu_B}}{e_0}).
\end{array}\\\tag{A23}
\end{equation}
where $e_d$ denotes errors from the channel relative-phase misalignment and $e_0=0.5$ is the error rate of a background noise.

Using the same method, we can also evaluate the overall gain and QBER for the triggered events and non-triggered events while taking the passive one intensity decoy-state protocol into account. For the photon number distribution of $P_{n}^{T}(\mu_x)$ and $P_{n}^{T}(\mu_x)$ ($x=A or B$), we can obtain the detection probabilities of four detectors by Eq.(\ref{HSPS-M}), and further evaluate the overall gain by Eq.(\ref{HSPS-S}) and QBER by Eq.(\ref{QBER-3}) and Eq.(\ref{QBER-10}). Taking the non-triggered events for an example, the probabilities of four detectors for detecting the non-triggered events are given by
\begin{equation}
\label{QBER-4}
\begin{array}{lll}
D^{NT}_{r_0}=D^{NT}_{s_0}=2a^{NT}_0p_d+P_{\mu_A}^{cor}\{\frac{v d_A(1-p_d)}{(1+x)(1+x\eta_D)}-\frac{d_Ax}{1+x}\\
\quad\quad\quad\quad\quad\quad+\frac{u}{z(z-u)}-\frac{w(1-p_d)}{z(z-w)}+\frac{d_Ap_d}{2(1+x)}-\frac{p_d}{2z}\},\\
D^{NT}_{r_1}=D^{NT}_{s_1}=P_{\mu_A}^{cor}\{\frac{v d_A(1-p_d)}{(1+x)(1+x\eta_D)}-\frac{d_A x}{1+x}\\
\quad\quad\quad\quad\quad\quad+\frac{u}{z(z-u)}-\frac{w(1-p_d)}{z(z-w)}-\frac{d_Ap_d}{2(1+x)}+\frac{p_d}{2z}\},
\end{array}\\\tag{A24}
\end{equation}
where $a^{NT}_0=\frac{1-P_{\mu_A}^{cor}}{2}+P_{\mu_A}^{cor}(\frac{1}{1+\mu_A(0.5\eta^c_A+\eta_A-0.5\eta_A\eta^c_A)}-\frac{d_A}{1+0.5\mu_A\eta^c_A})$. Then, in Eq.(\ref{HSPS-S}), replace $D_{r_0}$, $D_{s_0}$, $D_{r_1}$ and $D_{r_1}$ by $D^{NT}_{r_0}$, $D^{NT}_{s_0}$, $D^{NT}_{r_1}$ and $D^{NT}_{r_1}$ respectively and we can obtain $Q_{\mu_A\mu_B}^{(nt)}$. Similar to the calculation of $Q_{0\mu_B}$, $Q_{\mu_A0}$, and $E_{\mu_A\mu_B}$, we can also evaluate $Q_{0\mu_B}^{(nt)}$, $Q_{\mu_A0}^{(nt)}$, and $E_{\mu_A\mu_B}^{(nt)}$.

\section*{APPENDIX B: OVERALL GAIN AND QBER IN THE RECTILINEAR BASIS}
In appendix B, we will show how to calculate the overall gain and quantum bit error rate in the Z-basis, which is discussed in the paper by Ma et al\cite{MDI-d1}. In the rectilinear basis, both of the senders only choose one of the two $r$ and $s$ modes. Thus, the overall gain are divided into two parts: the gain $Q^{(C)}_Z$ from events that Alice and Bob choose the different modes, and the gain $Q^{(E)}_Z$  from errors that Alice and Bob choose the same mode and a successful two-click event occurs. Therefore, the overall gain in the Z-basis is given by
\begin{equation}
\label{AP-1}
\begin{array}{lll}
Q_Z=Q^{(C)}_Z+Q^{(E)}_Z.
\end{array}\\\tag{B1}
\end{equation}

In what follows, we shall discuss how to calculate $Q^{(C)}_Z$ and $Q^{(E)}_Z$. Since either Alice or Bob only choose one of the mode, the state fully passing the BS in sender's side without a transmittancy of $1/2$. Then, after passing through the channel, the state arrives at the BS in the relay has less $1/2$ times of channel transmittancy. And most importantly, now there exists no interference between two sender's states. So the gain $Q^{(C)}_Z$ is given by
\begin{equation}
\label{AP-2}
\begin{array}{lll}
Q^{(C)}_Z=2(1-D^{(Z_A)}_{r_1})(1-D^{(Z_B)}_{s_1})D^{(Z_A)}_{r_1}D^{(Z_B)}_{s_1},
\end{array}\\\tag{B2}
\end{equation}
where
\begin{equation}
\label{AP-3}
\begin{array}{lll}
D^{(Z_A)}_{r_1}=a_0p_d+\frac{P_{\mu_A}^{cor}}{P_{\mu_A}^{post}}\{\frac{x(1+d_A)}{1+x}-\frac{v(1+d_A)(1-p_d)}{(1+x)(1+x\eta_D)}\\
\quad\quad\quad\quad-\frac{u}{z(z-u)}+\frac{w(1-p_d)}{z(z-w)}\},\\
D^{(Z_B)}_{s_1}=b_0p_d+\frac{P_{\mu_B}^{cor}}{P_{\mu_B}^{post}}\{\frac{x(1+d_B)}{1+x}-\frac{v(1+d_B)(1-p_d)}{(1+x)(1+x\eta_D)}\\
\quad\quad\quad\quad-\frac{u}{z(z-u)}+\frac{w(1-p_d)}{z(z-w)}\}.
\end{array}\\\tag{B3}
\end{equation}

The calculation of gain from errors is slightly more complex. One of the error events is that Alice and Bob both choose $r$ mode. In this case, two photon interference occurs in the BS at the relay of $r$ mode. The detection probability of detector $r_0$ and $r_1$ can be calculated by Eq.(\ref{HSPS-M}). But the probability that one of the two detectors in the $s$ mode responses is given by $p_d(1-p_d)$. Thus, the formula for calculating $Q^{(E)}_Z$ is shown as the following
\begin{equation}
\label{AP-4}
\begin{array}{lll}
Q^{(E)}_Z=2p_d(1-p_d)D^{(Z')}_{r_0}(1-D^{(Z')}_{r_1}),
\end{array}\\\tag{B4}
\end{equation}
where
\begin{equation}
\label{AP-5}
\begin{array}{lll}
D^{(Z')}_{r_0}=2a'_0p_d+\frac{P_{\mu_A}^{cor}}{P_{\mu_A}^{post}}\{\frac{x'(1+d_A)}{1+x'}-\frac{v'(1+d_A)(1-p_d)}{(1+x')(1+x'\eta_D)}\\
\quad\quad\quad\quad-\frac{u'}{z'(z'-u')}+\frac{w'(1-p_d)}{z'(z'-w')}-\frac{(1+d_A)p_d}{2(1+x')}+\frac{p_d}{2z'}\},\\
D^{(Z')}_{r_1}=\frac{P_{\mu_A}^{cor}}{P_{\mu_A}^{post}}\{\frac{x'(1+d_A)}{1+x'}-\frac{v'(1+d_A)(1-p_d)}{(1+x')(1+x'\eta_D)}\\
\quad\quad\quad\quad-\frac{u'}{z'(z'-u')}+\frac{w'(1-p_d)}{z'(z'-w')}+\frac{(1+d_A)p_d}{2(1+x')}-\frac{p_d}{2z'}\},
\end{array}\\\tag{B5}
\end{equation}
with $a'_0=1-P_{\mu_A}^{cor}+\frac{P_{\mu_A}^{cor}}{P_{\mu_A}^{post}}(\frac{1+d_A}{1+\mu_A\eta^c_A}-\frac{1}{1+\mu_A(\eta^c_A+\eta_A-\eta_A\eta^c_A)})$, $x'=\mu_{A}\eta_{A}^{c}$, $z'=1+x'+\mu_A\eta_A-x'\eta_A$, $u'=x'(1-\eta_A)$, $v'=x'(1-\eta_D)$ and $w'=u'(1-\eta_D)=v'(1-\eta_A)$.

For the calculation of quantum bit error rate, taking additional misalignment errors into account, we obtain
\begin{equation}
\label{AP-6}
\begin{array}{lll}
E_Z Q_Z=e_d Q^{(C)}_Z+(1-e_d)Q^{(E)}_Z.
\end{array}\\\tag{B6}
\end{equation}

 The overall gain and QBER under Z-basis for the triggered events and non-triggered events in the passive one intensity decoy-stated protocol can be obtained in the same way. However, one should consider the photon number distribution shown as Eq.(\ref{Passive-01},\ref{Passive-02}) other than Eq.(\ref{HSPS-C}).

\end{document}